\documentclass[a4paper,11pt]{article}
\pdfoutput=1 

\usepackage{jcappub} 

\usepackage[T1]{fontenc} 

\title{\boldmath The cosmological collider in $R^2$ inflation}

\author[a]{Yi-Peng Wu}

\affiliation[a]{Institute of Physics, Academia Sinica, Taipei 115201, Taiwan}

\emailAdd{ypwu@gate.sinica.edu.tw}

\abstract{
Starobinsky's $R^2$ inflation manifests a best-fit scenario for the power spectrum of primordial density fluctuations. Observables derived from the slow-roll picture of the $R^2$ model in the Einstein frame relies on the conformal transformation of the metric, which inevitably induces a unique exponential-type couplings of the rolling scalaron with all matter fields during inflation. The ``large-field'' nature of the $R^2$ model further invokes non-negligible time and scale dependence to the matter sector through such an exponential coupling, modifying not only the dynamics of matter perturbations on superhorizon scales but also their decay rates. In this work, we identify the simplest observable of the cosmological collider physics built in the background of $R^2$ inflation, focusing on the so-called ``quantum primordial clock'' signals created by the non-local propagation of massive scalar perturbations. Our numerical formalism based on the unique conformal coupling can have extended applications to (quasi-)single-field inflationary models with non-trivial couplings to gravity or models that originated from the $f(R)$ modification of gravity.
}

\definecolor{linkcolor}{RGB}{41, 127, 255}
\hypersetup{
	pdfstartview={XYZ},
	colorlinks,
	allcolors=linkcolor,
	bookmarksopen
}

\begin{document}
	\maketitle
	\flushbottom

\section{Introduction}\label{Sec. introduction}
\textit{Planck} measurements of the primordial scalar power spectrum provide precision tests of the single-field inflationary models within the framework of Einstein gravity \cite{Planck:2013jfk,Planck:2015sxf,Planck:2018jri}. In the latest report \cite{Planck:2018jri}, the power spectrum of the curvature perturbation with a spectral index $n_s = 1$  has been ruled out by more than $6\sigma$, reinforcing the slow-roll paradigm of single-field inflation with small deviations from exact scale invariance. Up to $95 \%$ confidence level, the observational data favor slow-roll models with concave inflaton potentials, and among representative theories, Starobinsky's $R^2$ model \cite{Starobinsky:1980te} manifests the best-fit predictions for the central value of the $n_s$ distribution and the bounded tensor-to-scalar ratio $r$. 
 
Next-generation experiments (such as Simons Observatory \cite{SimonsObservatory:2018koc}, LiteBIRD \cite{LiteBIRD:2020khw} and CMB-S4 \cite{CMB-S4:2016ple}) will aim to improve the current sensitivity bound on $r$ to be at least one-order-of-magnitude finer. However, due to the high degeneracy in the predictions from all kinds of viable models, 
the importance of finding observables to test inflationary theories beyond the joint $n_s$-$r$ constraints has been stressed \cite{Braglia:2022ftm}.
Recently, a possibility of testing large-field inflation by virtue of the well-established cosmological collider physics \cite{Chen:2009we,Chen:2009zp,Arkani-Hamed:2015bza} (see also \cite{Dimastrogiovanni:2015pla,Schmidt:2015xka,Kehagias:2015jha,Meerburg:2016zdz,Lee:2016vti,Chen:2016hrz,Chen:2016uwp,Kehagias:2017cym,Franciolini:2017ktv,Kumar:2017ecc,MoradinezhadDizgah:2018ssw,Saito:2018omt,Kumar:2018jxz,Goon:2018fyu,Wu:2018lmx,Lu:2019tjj,Liu:2019fag,Lu:2021wxu,Cui:2021iie,Tong:2022cdz,Reece:2022soh,Qin:2022lva,Bodas:2020yho,Wang:2019gbi,Wang:2020ioa,McCulloch:2024hiz,Yin:2023jlv,Jazayeri:2023xcj,Chen:2022vzh,Aoki:2023tjm,Craig:2024qgy,Cabass:2024wob,Ema:2023dxm,Sohn:2024xzd,Aoki:2024uyi,Werth:2023pfl,Pinol:2023oux,Werth:2024aui,Aoki:2020zbj,Pinol:2021aun}) has been pointed out in \cite{Reece:2022soh}.
In this work, we further dig into the family of large-field models, focusing on the simplest observable in the cosmological collider for the best-fitting $R^2$ inflation.
This simplest observable, sometimes is referred as the ``quantum primordial standard clock \cite{Chen:2015lza,Domenech:2018bnf,Domenech:2020qay}'' in the ideal cosmological collider, exhibits the most symbolic prediction from the heavy-mass regime of quasi-single-field inflation \cite{Chen:2009we,Chen:2009zp,Baumann:2011nk,Chen:2012ge,Pi:2012gf,Assassi:2013gxa,Noumi:2012vr,Gong:2013sma,An:2017hlx,An:2017rwo,Tong:2017iat,Iyer:2017qzw,Wang:2018tbf,McAllister:2012am,Garcia-Saenz:2019njm}.

The ideal cosmological collider is built with approximately exact isometries of the de Sitter spacetime where the analytic structure of correlation functions for the curvature perturbation can be ``bootstrapped'' from the final boundary surface at the end of inflation without even knowing the detailed time evolution during the slow-roll phase \cite{Arkani-Hamed:2018kmz,Pimentel:2022fsc,Jazayeri:2022kjy,Qin:2022fbv,Xianyu:2022jwk,Wang:2022eop,Qin:2023ejc,Arkani-Hamed:2023bsv,Arkani-Hamed:2023kig,Baumann:2019oyu,Baumann:2020dch,Sleight:2019hfp,Pajer:2020wxk,Goodhew:2020hob}. In this sense, standard predictions from the cosmological collider physics, typically referring to oscillatory momentum scaling in the spectra of primordial correlation functions created by massive particle productions, are independent of the model that creates the inflationary background. These oscillatory features characterize the non-local propagation of massive particles during inflation, which cannot be mimicked by any effective field theory of the inflaton alone, and they are usually exponentially suppressed and can only exist in three-point or higher-order correlation functions. However, due to the explicit analytic expressions, the standard predictions of the cosmological collider may serve as useful baselines to measure the model-dependent signals from realistic inflationary models.

Realistic slow-roll inflation must at the minimum violate the de Sitter dilational invariance \cite{Antoniadis:2011ib} (or the dilatation symmetry) between space and time due to the time evolution of inflaton $\phi = \phi(t)$. However, the size of such a dilatation symmetry breaking is suppressed by the smallness of slow-roll parameters and it is in general not easy to result in important observational consequences to the spectra of primordial correlators.  For large-field inflation with an inflaton excursion range larger than the Planck scale ($\Delta\phi \gtrsim M_P$), the pioneer study \cite{Reece:2022soh} has identified the effect of dilatation violation in the cosmological collider observables from the scalar particle production with a time-dependent mass. This time-varying effect could be induced by the exponential type coupling $\sim e^{-\alpha \phi/M_P}$ with the slow-rolling inflaton $\phi$ due to the super-Planckian completion of the inflationary models, and in this case the time-varying effect is only suppressed by the square root of the slow-roll parameter ($\sqrt{\epsilon} \sim \dot{\phi}/M_PH$) and it modifies both the spectral amplitude and oscillatory frequencies from the standard predictions with a constant mass. Recently, the analytic expression of primordial spectra featuring such a dynamical mass effect with couplings in the limit of $\alpha \ll 1$ is found in \cite{Aoki:2020wzu}.

$R^2$ model is essentially a large-field inflation. However, without additional assumptions, the exponential coupling $\sim e^{-\alpha \phi/M_P}$ with all kinds of matter fields  in $R^2$ inflation naturally arises due to the conformal transformation of the metric into the so-called Einstein frame and $\alpha = \sqrt{2/3}$ is uniquely fixed by the canonical normalization of inflaton in the Einstein frame. Moreover, such an exponential coupling introduced by the metric conformal transformation (which we shall refer as ``conformal coupling'' for short) not only acts on the particle masses (or the potentials) but also attaches to the kinetic terms. This can lead to different dilution behavior of the quantum fluctuations in the inflationary background.

The study of the cosmological collider physics in $R^2$ inflation can have more implications to modified gravity theories beyond the Einstein-Hilbert action and for single-field inflationary models with non-trivial coupling with gravity:
\begin{itemize}
	\item The $R^2$ model is a special case of the $f(R)$ gravity, while the $f(R)$ gravity can be cast into a generalized version of the Brans-Dicke theory (or more generally the scalar-tensor theory) \cite{Sotiriou:2008rp,DeFelice:2010aj,Nojiri:2010wj,Nojiri:2017ncd}. The Einstein frame of any $f(R)$ gravity theory possesses the same type of conformal coupling with matter fields as studied in this work (see Appendix~\ref{Append_R2}).
	\footnote{This statement is based on the metric formalism.} 
	
	\item The Einstein frame of the $R^2$ model is a useful representation to investigate the slow-roll dynamics of inflation. There has been open debates on the equivalence between the Jordan frame and Einstein frame representations. Since the (conformal) coupling of matter with the slow-roll inflaton (namely the scalaron) does not trivially exist in the Jordan frame, we take the calculation in this work for granted as predictions based on the Einstein frame.
	
	\item The inflaton potential of the $R^2$ model in the Einstein frame coincides with the specific limits of many other inflationary model potentials with more free parameters. Some of those potentials may also be converted via the conformal transformation from their original frame, and this shall generate a similar coupling with matter fields as that in $R^2$ inflation. In particular,  the large-field expansion of the conformal coupling in the Higgs inflation with a non-minimal coupling to gravity \cite{Bezrukov:2007ep} takes exactly the same form as that of the $R^2$ inflation at the leading order (see Appendix~\ref{Appen_Higgs_inflation}).
\end{itemize}

This work is organized as follows: 

In Section~\ref{Sec_mode_functions}, we first review the conformal coupling in the Einstein frame of the $R^2$ inflation and we verify the corresponding interaction vertices induced by such a conformal coupling in Section~\ref{Sec_R2_background}. We provide two numerical methods to resolve the mode functions of a scalar perturbation in Section~\ref{Sec_constant_epsilon} \& \ref{Sec_SR_scalaron}, and we compare these solutions with an analytic formalism for a time-varying mass in Section~\ref{Sec_dynamical_mass}.

We study the leading corrections to the scalaron power spectrum in $R^2$ inflation generated by the transfer of massive scalar perturbations in Section~\ref{Sec_Power_Spectrum}. In Section~\ref{Sec_Bispectrum}, we clarify the simplest observable of the cosmological collider physics, namely the quantum primordial clock signals from the inflationary background created by the $R^2$ model. Finally, conclusions and discussions of the main findings in this work are provided in Section~\ref{Sec. conclusion}. 

\section{Scalar mode functions} \label{Sec_mode_functions}

\begin{figure}[]
	\begin{center}
		\includegraphics[width= 6 cm]{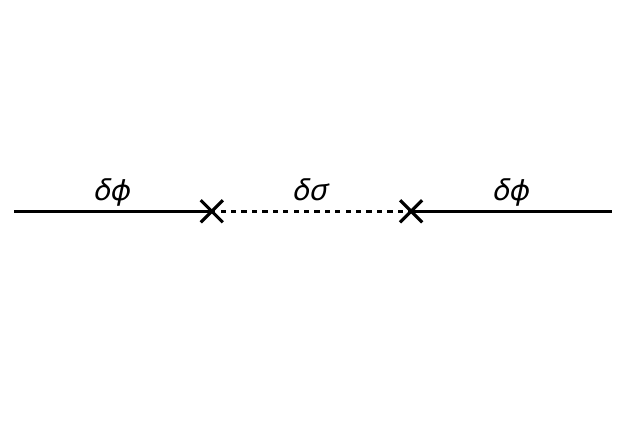}
		\hfill
		\includegraphics[width= 6 cm]{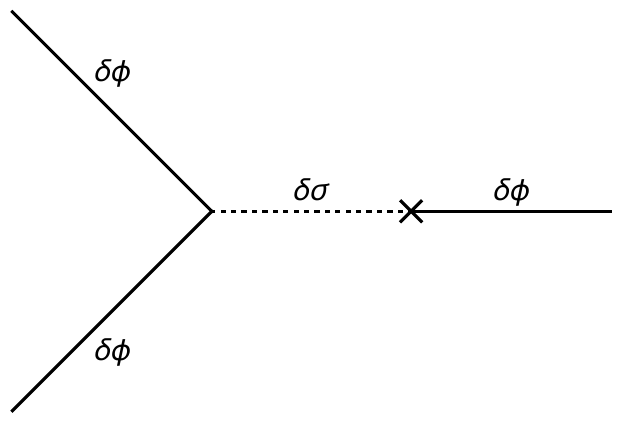}
	\end{center}
	\caption{\label{fig.Feynman_clock} 
		Diagrammatic illustration of the correction to the power spectrum (Left Panel) computed in Section~\ref{Sec_Power_Spectrum} and the corresponding three-point correlation for the simplest quantum primordial clock (Right Panel) in Section~\ref{Sec_Bispectrum}. 
	}
\end{figure}

In this work we aim to work out the correlation functions of the curvature perturbation $\zeta$ involved with the exchange of a scalar particle $\sigma$ as illustrated in Figure~\ref{fig.Feynman_clock}. The physical picture is clear if we restrict ourselves in the spatially flat gauge so that the curvature perturbation is related to the linear inflaton perturbation via $\zeta = -H \delta\phi / \dot{\phi}_0$, where $\phi_0(t)$ describes the homogeneous (or the zero-mode) evolution of the inflaton. This simple relation relies on the fact that $H$ and $\dot{\phi}_0$ are nearly constants during inflation. Indeed, this is a good approximation for slow-roll inflationary models with the first slow-roll parameter $\epsilon_1 \equiv -\dot{H}/H^2 \ll 1$. Currently we have a conservative upper bound $\epsilon_1 < 0.0063$ from \cite{Arkani-Hamed:2018kmz}, and the leading corrections to $H$ and $\dot{\phi}_0 \approx \sqrt{2\epsilon_1} M_P H$ are at least suppressed by factors of $\mathcal{O}(\epsilon_1)$. This means that important time corrections to $H$ or $\dot{\phi}_0$ need to be accumulated over a number of $e$-folds $\Delta N > 30$ from the pivot scale $k_\ast = 0.002$ Mpc${}^{-1}$ of observations \cite{Arkani-Hamed:2018kmz}. However, having in mind that the practical experiments could only scan a range of scales across $\Delta N \sim \mathcal{O}(10)$, we will treat $H$ and $\dot{\phi}_0$ as constants in time throughout the current investigation.

As we will see, interactions with the inflaton $\phi$ in the $R^2$ inflation are typically originated from gravitational effects and thus they all suppressed by the (reduced) Planck scale $M_P \approx 2.4 \times 10^{18}$ GeV. This ensures that the system considered in this work never enters the ``strong-coupling'' regime and we can adopt the standard analytic mode functions in the inflationary background for the ``free-field'' inflaton perturbations.    

\subsection{The $R^2$ inflationary background}\label{Sec_R2_background}
Let us start with a brief review of the slow-roll picture in the $R^2$ inflation model. More details related to the conformal transformation of the metric can be found in Appendix~\ref{Appen_A}.
The action relevant to our study is given by:
\begin{align}
\label{Action_J_frame}
	S &= \int d^4 x \sqrt{-g_J}\, \frac{M_P^2}{2} \left[R_J +\frac{R_J^2}{6M^2}\right] &&+\int d^4x \sqrt{-g_{J}}\, \mathcal{L}_{\rm matter}, 
\\\label{Action_E_frame}
	   &= \int d^4x \sqrt{-g_E}\, \left[\frac{M_P^2}{2} R_E - \frac{1}{2}(\partial\phi)_E^2 -U(\phi)\right] &&+
	 \int d^4x \sqrt{-g_E}\, e^{-2\sqrt{\frac{2}{3}}\frac{\phi}{M_P}} \mathcal{L}_{\rm matter},
\end{align}
where $R_J$ ($R_E$) is the Ricci scalar in the Jordan (Einstein) frame constructed via the metric $g_{\mu\nu}^J = e^{-\sqrt{\frac{2}{3}}\frac{\phi}{M_P}} g_{\mu\nu}^E$ and the definition of the conformal factor $\Omega^2 = e^{\sqrt{\frac{2}{3}}\frac{\phi}{M_P}}$ makes sure that the inflaton $\phi$ is canonically normalized in the Einstein frame \eqref{Action_E_frame}. Note that $(\partial\phi)_E^2 \equiv g_E^{\mu\nu} \partial_\mu\phi\partial_\nu\phi$. The mass scale $M$ controls the Hubble scale of inflation $H$ in the Einstein frame and $\mathcal{L}_{\rm matter}$ can include all kinds of matter fields during inflation. $\phi$ is also known as the ``scalaron,'' which can be viewed as the longitudinal degree of freedom of the gravitational sector in the higher-order extension of the Einstein's field equation \cite{DeFelice:2010aj,Sotiriou:2008rp}.

 It is straightforward to compute observables of the $R^2$ inflation in the Einstein frame based on the scalaron potential
\begin{align}\label{U_phi}
	U(\phi) = \frac{3}{4} M_P^2M^2 \left(1-  e^{-\sqrt{\frac{2}{3}}\frac{\phi}{M_P}}\right)^2.
\end{align}
The slow-roll parameters of the $R^2$ model can be obtained from $U(\phi)$. For example, the first slow-roll parameter reads
\begin{align}
	\epsilon_U(\phi) = \frac{M_P^2}{2} \left(\frac{U_\phi}{U}\right)^2 = \frac{4}{3} \left(e^{\sqrt{\frac{2}{3}}\frac{\phi}{M_P}} -1 \right)^{-2}.
\end{align}
One can check that the plateau region of the potential $U(\phi)$ for realizing the slow-roll dynamics ($\epsilon_U \ll 1$) is in the limit of $\phi/M_P \gg 1$, which implies that the $R^2$ model should be cast into the ``large-field'' inflation scenario.  For a given epoch in terms of the $e$-folding number $N \equiv \ln a$ during inflation, the corresponding scalaron value can be obtained via the formula
\begin{align}\label{def_e_fold_phi}
	\frac{\phi}{M_P} \approx \sqrt{\frac{3}{2}} \ln \left(\frac{4}{3}\; \Delta N\right),
\end{align}
where $\Delta N = N_{\rm end} - N \geq 0$ is the number of $e$-folds from the given epoch to the end of inflation. 
See Figure~\ref{fig.slowroll_scalaron} for a comparsion of \eqref{def_e_fold_phi} with the numerical solution of $\phi(N)$.
This formula allows us to express $\epsilon_U$ as a function of $\Delta N$.

The most important thing to be outlined here is that: as long as we work out the slow-roll dynamics of the $R^2$ model in the Einstein frame \eqref{Action_E_frame}, the matter sector inevitably receives a universal exponential-type coupling with the inflaton $\phi$ invoked by the conformal tranformation of the metric. As pointed out in \cite{Reece:2022soh}, large-field inflation with such an exponential-type coupling can introduce mild, yet non-negligible, scale dependence to the equation of motion of a matter field, modifying solutions of the quantum mode fluctuations from the standard cases with the dilatation symmetry.

To see more explicitly, let us consider as a typical example, $\mathcal{L}_{\rm matter} = -\frac{1}{2}(\partial\sigma)_J^2 -V(\sigma)$ is a canonical scalar in the Jordan frame \eqref{Action_J_frame}, where $(\partial\sigma)_J^2 = e^{\sqrt{\frac{2}{3}}\frac{\phi}{M_P}} g^{\mu\nu}_E\partial_\mu\sigma\partial_\nu\sigma$ after the conformal transformation. 
In the Einstein frame \eqref{Action_E_frame}, the equation of motion for the linear perturbation of the isocurvature scalar $\sigma = \sigma_0(t) + \delta\sigma(t, \vec{x})$ becomes
\begin{align}\label{eom_sigma_general}
	\ddot{\delta\sigma} + \left(3 - \sqrt{\frac{2}{3}} \frac{\dot{\phi}_0}{M_PH}\right)H \dot{\delta\sigma} 
	+ \left(\frac{k^2}{a^2} +e^{-\sqrt{\frac{2}{3}}\frac{\phi_0}{M_P}} V_{\sigma\sigma} \right) \delta\sigma =0,
\end{align} 
where $V_{\sigma\sigma} = \partial_\sigma^2 V(\sigma)$ describes the effective mass around a local minimum at a non-trivial vacuum expectation value (VEV) $\sigma_0 \neq 0$ such that $\partial_\sigma V\vert_{\sigma = \sigma_0} = 0$.
Metric perturbations can be omitted in this equation of motion since the energy density of $\sigma$ only contributes a negligible fraction to the total. Comparing \eqref{eom_sigma_general} to the case of large-field inflation in \cite{Reece:2022soh}, one can see that in $R^2$ inflation the conformal transformation introduces a same time-dependent factor $\sim e^{-\alpha \phi_0(t)/M_P}$ to the mass term of $\delta\sigma$ with $\alpha = \sqrt{2/3}$. Moreover, the conformal coupling also introduces extra modifications to the Hubble friction term.

If we restrict to the tree-level processes given in Figure~\ref{fig.Feynman_clock}, the transfer vertices or the interactions vanishes identically from the coupling with potential $\sim e^{-2\alpha\phi_0/M_P}V(\sigma)$, since we consider a non-trivial VEV $\sigma_0$ near a local minimum with $V_\sigma = \partial_\sigma V= 0$. This also makes cubic vertices arising from derivative of the scalar potential, such as $V_{\sigma\phi\phi}$, vanishing.
Thus we shall focus on the coupling from the kinetic term. 
Note that the scalaron $\phi$ may involve in the scalar potential in the Einstein frame if $\sigma$ has non-trivial couplings with gravity in the Jordan frame. In such cases $\sigma$ will become a part of the scalaron under conformal transformation.
\footnote{The mixed Higgs-$R^2$ inflation \cite{He:2018gyf,Ema:2023dxm} provides an example of such a case by including a non-minimal coupling with gravity $\sim \xi \sigma^2 R_J $.  See also Appendix~\ref{Appen_mixed_Higgs_R2} for more details.
	}
  
In the simplest case with a static VEV $\dot{\sigma}_0 = 0$, the two-point transfer vertex for the scalar perturbations $\delta\sigma$ and $\delta\phi$ in Figure~\ref{fig.Feynman_clock} vanishes from the kinetic term $\sim e^{-\alpha\phi_0/M_P}(\partial\sigma)_E^2$ and the leading contribution will come from loop diagrams. To allow the presence of the two-point vertex $\sim e^{-\alpha\phi_0/M_P}\dot{\sigma}_0\delta\phi \delta\dot{\sigma}$, we assume a zero-mode motion $\dot{\sigma}_0 \neq 0$ around the pivot scale of our interest, where $\dot{\sigma}_0$ must be small enough to preserve a good approximation of the scalaron solution \eqref{slow_roll_phi_analytic} in the single-field inflation limit. A possible realization of these conditions is given in Appendix~\ref{Appen_mixed_Higgs_R2}. 

In order to examine our numerical computations with the known analytic results in previous studies, we perform  integration by parts to the conformal coupling with the kinetic term in \eqref{Action_E_frame} to obtain the following vertices and interactions:
\begin{align}
	\label{def_L2}
	\delta\mathcal{L}_2 &\subset \frac{c_2}{M_P^2} a^3 e^{-\sqrt{\frac{2}{3}}\frac{\phi_0}{M_P}} \sigma_0\dot{\phi}_0 \delta\dot{\phi}\delta\sigma,\\
	\label{def_L3}
	\delta\mathcal{L}_3 &\subset \frac{c_3}{M_P^2} a^3 e^{-\sqrt{\frac{2}{3}}\frac{\phi_0}{M_P}} \sigma_0 \delta\sigma (\partial\delta\phi)^2,
\end{align}
where $c_2$, $c_3$ are constant values of $\mathcal{O}(1)$. We will use \eqref{def_L2} and \eqref{def_L3} for the calculations of the processes  given in Figure~\ref{fig.Feynman_clock}, which is more convenient for extracting the effects led by the conformal coupling. 
 In the limit of $\dot{\sigma}_0 \rightarrow 0$, the overall tree-level contribution from the original kinetic term $\sim e^{-\alpha\phi_0/M_P}(\partial\sigma)_E^2$ goes away, which means that the contribution from  \eqref{def_L2} and \eqref{def_L3} is exactly cancelled out by the other terms from the integration by parts.
Thus, with $\dot{\sigma}_0 \neq 0$, the contribution from the original vertices is expected to be the results based on \eqref{def_L2} and \eqref{def_L3} times a suppression factor given by the smallness of the non-vanished $\dot{\sigma}_0$. This suppression factor depends on the explicit model for the isocurvature field. (It is the non-minimal coupling \eqref{suppression_factor_xi} if we consider the model of Appendix~\ref{Appen_mixed_Higgs_R2}.)

Note that interactions with $\phi$ originated from the conformal factor are always suppressed by $M_P$, since the scalaron is the longitudinal mode of gravity. This also ensures that mode functions solved by \eqref{eom_sigma_general} are indeed ``free-field'' solutions in the interaction picture. 

\subsection{The near pivot-scale expansion}\label{Sec_constant_epsilon}
\begin{figure}[]
	\begin{center}
		\includegraphics[width=7 cm]{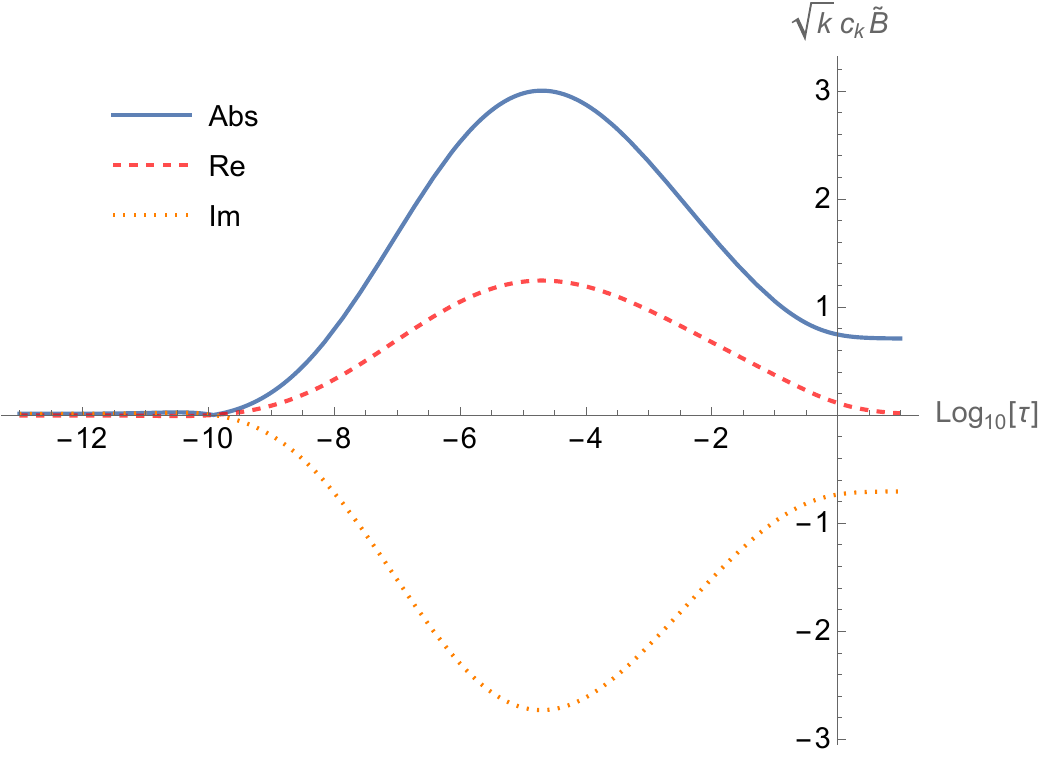}
		\hfill
		\includegraphics[width= 7 cm]{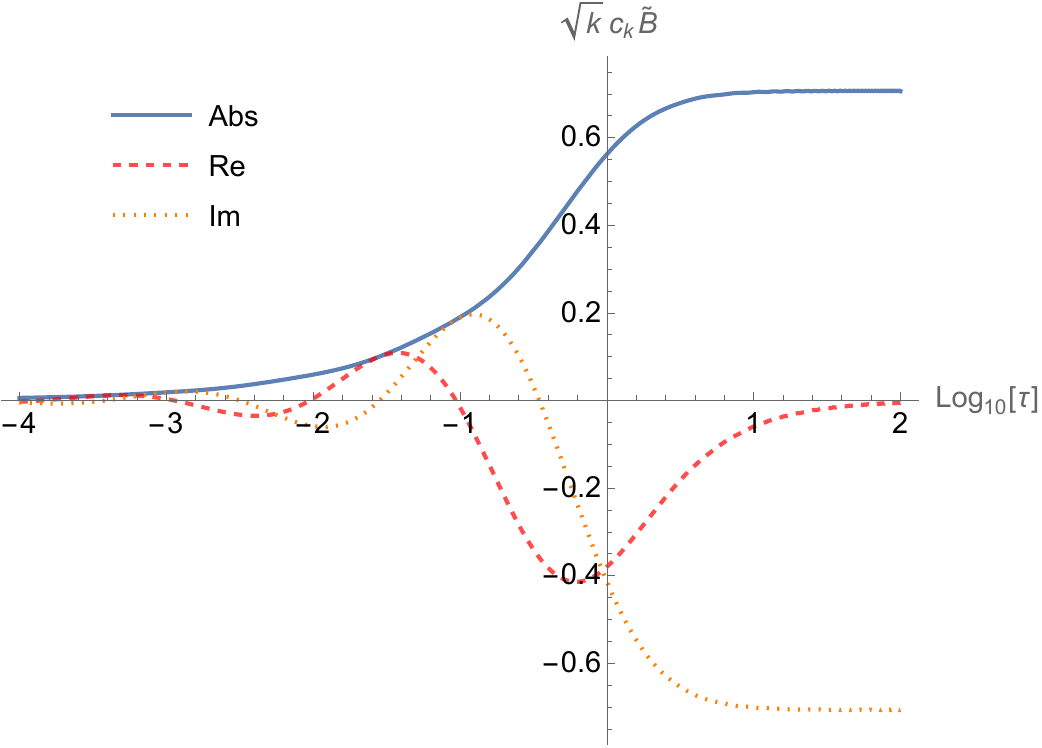}
	\end{center}
	\caption{\label{fig.B_evolution} 
		The evolution of the factorized mode function $\tilde{B}$ with respected to $\tau = -k\eta$ for $M_\ast = 1.3$ (Left Panel) and $M_\ast = 2$ (Right Panel). Here we use $k = k_\ast = 0.002$ Mpc${}^{-1}$, and a constant $\Delta_\epsilon = \Delta_{55}$. 
	}
\end{figure}

We now solve \eqref{eom_sigma_general} by expanding the potential $V(\sigma)$ around its stable VEV $\sigma =\sigma_0$ to obtain a constant mass parameter $V_{\sigma\sigma} = m_\sigma^2$. 
Following the approach used in \cite{Reece:2022soh}, we adopt the expansion around the pivot scale $k_\ast$ for observations:
\begin{align}
	\phi_0(t) = \phi_0(t_\ast) - \vert\dot{\phi}(t_\ast)\vert (t-t_\ast),
\end{align}
where $t_\ast$ (or $\eta_\ast$) is the (conformal) time when the pivot scale $k_\ast$ crosses the horizon at which $-k_\ast \eta_\ast = 1$. The valid range for this expansion is $k_{\rm min} \ll k \ll k_{\rm max}$, with $k_{\rm min} = e^{-10}k_\ast$ and $k_{\rm max} = e^{10}k_\ast$. Note that $\dot{\phi}_0 < 0$ in $R^2$ inflation.

As we get rid of the mild time dependence in $H$ and $\dot{\phi}_0$, we obtain a constant slow-roll parameter $\epsilon_U(t_\ast) = \dot{\phi}_0^2(t_\ast)/2M_P^2H^2(t_\ast)$. Now the non-negligible time dependence shows only in the exponential conformal factor, as
\begin{align}
	e^{-\sqrt{\frac{2}{3}}\frac{\phi_0}{M_P}} = e^{-\sqrt{\frac{2}{3}}\frac{\phi_\ast}{M_P}} \left(\frac{a}{a_\ast}\right)^{\sqrt{\frac{2}{3}} \sqrt{2\epsilon_U(t_\ast)}},
\end{align}
where $\phi_\ast = \phi(t_\ast)$ and $a_\ast = a(t_\ast)$.
As a result, we can rewrite \eqref{eom_sigma_general} into the form of
\begin{align}\label{eom_sigma}
	\ddot{\delta\sigma} + \left(3 + \Delta_\epsilon\right) H \dot{\delta\sigma} +
	\left[\frac{k^2}{a^2} + \left(\frac{a}{a_\ast}\right)^{\Delta_\epsilon} m_\ast^2\right] \delta\sigma = 0,
\end{align}
with $ m_\ast^2 =  e^{-\sqrt{2/3} \phi_\ast/M_P} m_\sigma^2$. The parameter $\Delta_\epsilon = \sqrt{\frac{2}{3}} \sqrt{2\epsilon_U}$ measures the departure from the exactly dilatation invariant case. Namely, the standard mode function for a massive scalar shall be reproduced in the limit of $\Delta_\epsilon\rightarrow 0$.

The constant $\epsilon_U$ (and therefore a constant $\Delta_\epsilon$) expansion used in this section allows us to proceed the further transformation:
\begin{align}
	\delta\sigma = a^{-1-\Delta_\epsilon/2} \delta\tilde{\sigma} = (-H\eta)^{1+\Delta_\epsilon/2} \delta\tilde{\sigma}.
\end{align}
Here we consider the canonical quantization of $\delta\tilde{\sigma}$ as
\begin{align}
	\delta\tilde{\sigma}(\eta,\vec{x}) = \int\frac{d^3\vec{k}}{(2\pi)^3} e^{i\vec{k}\cdot\vec{x}}
	\left[\tilde{u}_k(\eta) \hat{b}_{\vec{k}} + \tilde{u}_k^\ast(\eta) \hat{b}_{-\vec{k}}^\dagger\right],
\end{align}
with $[\hat{b}_{\vec{k}},\hat{b}_{-\vec{p}}^\dagger] = (2\pi)^3 \delta^3(\vec{k}+\vec{p})$. 
In terms of the dimensionless time parameter $\tau \equiv -k\eta$, which is more convenient for numerical computations, we further translate the equation \eqref{eom_sigma} into
\begin{align}\label{eom_tilde_u}
	\frac{\partial^2}{\partial\tau^2} \tilde{u}_k +
	\left[1-\left(1+\frac{\Delta_\epsilon}{2}\right)\left(2+\frac{\Delta_\epsilon}{2}\right)\frac{1}{\tau^2} + \frac{M_\ast^2}{\tau^{2+\Delta_\epsilon}} \left(\frac{k}{k_\ast}\right)^{\Delta_\epsilon}\right] \tilde{u}_k =0,
\end{align}
where $M_\ast^2 \equiv m_\ast^2/H^2$ is the dimensionless mass parameter.

One can see that the differential equation \eqref{eom_tilde_u} becomes inhomogeneous if $\Delta_\epsilon \neq 0$, and thus the mode function $\tilde{u}_k$ in general can only be solved numerically. Due to the dilatation symmetry of the de Sitter space, the formulation in terms of the dimensionless time $\tau = -k\eta$ translates the time dependence in the conformal coupling of \eqref{eom_sigma_general} to be explicit $k$-dependence. 

If somehow we could turn off $\Delta_\epsilon$ only to the mass term in \eqref{eom_tilde_u}, the equation of motion becomes
\begin{align}\label{eom_tilde_u_modified_d}
	\frac{\partial^2}{\partial\tau^2} \tilde{u}_k +
	\left[1-\left(1+\frac{\Delta_\epsilon}{2}\right)\left(2+\frac{\Delta_\epsilon}{2}\right)\frac{1}{\tau^2} + \frac{M_\ast^2}{\tau^{2}} \right] \tilde{u}_k =0.
\end{align}
This artifical equation exhibits the analytic solution as
\begin{align}\label{def_nu_d}
	\tilde{u}_k = c_k(\nu) \sqrt{\tau} H_{\nu}^{(1)}(\tau), \; \quad
	\nu(\Delta_\epsilon,M_\ast) = \sqrt{\left(\frac{3+\Delta_\epsilon}{2}\right)^2 - M_\ast^2},
\end{align}
which recovers the correct Bunch-Davies vacuum state, $\tilde{u}_k \rightarrow -i e^{i\tau}/\sqrt{2k}$, in the early-time limit with $\tau \rightarrow\infty$. It provides useful information for us to discover the initial conditions for our true solutions.
\footnote{We use $\nu = \sqrt{(\frac{3+\Delta_\epsilon}{2})^2 - M_\ast^2}$ for $0 \leq M_\ast < 3/2$ and $\mu=-i\nu = \sqrt{M_\ast^2 -  ( \frac{3+\Delta_\epsilon}{2} )^2}$ for $M_\ast \geq 3/2$.}

Let us return to the realistic situation of \eqref{eom_tilde_u}. Since we will need to deal with a convolution of time in the correlation functions of the inflaton, for a better numerical performance, we propose a factorization of the mode function as
\begin{align}\label{def_B_tilde}
	\tilde{u}_k(\tau) = c_k(\nu) \tilde{B}(\tau) e^{i \tau}, \qquad c_k(\nu) = -\frac{i}{2}\sqrt{\frac{\pi}{k}} e^{i(\nu+1/2)\pi/2},
\end{align}
where $\nu = \nu(\Delta_\epsilon, M_\ast)$ is defined in \eqref{def_nu_d}.
Such a factorization is motivated by the equation-of-motion approach \cite{Chen:2015dga} for the strongly coupled regime in the quasi-single-field inflation \cite{An:2017hlx,An:2017rwo,Wang:2018tbf}. Remarkably, we only adopt the factorization to the mode function $\tilde{u}_k$ which has to be solved numerically, while, on the other hand, our system is weakly coupled so that the inflaton mode function takes the standard analytic form of a massless scalar in the de Sitter space.

In terms of the factorized mode function $\tilde{B}(k, \tau)$, we finally arrive at the equation of motion reads
\begin{align}\label{eom_tilde_B}
	\frac{\partial^2}{\partial\tau^2} \tilde{B} +2i \frac{\partial}{\partial\tau} \tilde{B} +
	\left[ \frac{M_\ast^2}{\tau^{2+\Delta_\epsilon}} \left(\frac{k}{k_\ast}\right)^{\Delta_\epsilon} - \left(1+\frac{\Delta_\epsilon}{2}\right)\left(2+\frac{\Delta_\epsilon}{2}\right)\frac{1}{\tau^2}  \right] \tilde{B} = 0.
\end{align}
A good news is that the mode function $\tilde{u}_k$ (or $\tilde{B}$) does not recognize the broken dilatation invariance led by a non-zero $\Delta_\epsilon$ in the flat-space (or namely the early-time) limit with $\tau \rightarrow\infty$. This means that the mode function shares the standard Bunch-Davies vacuum state, $\tilde{u}_k \rightarrow -i e^{i\tau}/\sqrt{2k}$, with the case of $\Delta_\epsilon = 0$. By matching the analytic solution of a standard massive scalar in the vacuum (see Appendix~\ref{Appen_B} for more details), we obtain the initial conditions in the limit of $\tau \rightarrow\infty$ as
\begin{align}\label{def_B_tilde_initial_condition}
	\tilde{B} = \sqrt{\frac{2}{\pi}} e^{-i\pi \nu/2} e^{-i\pi/4}, \qquad 
	\frac{\partial}{\partial\tau} \tilde{B} = \sqrt{\frac{2}{\pi}} e^{-i\pi \nu/2} e^{-i\pi/4} \left(\frac{1}{2}-\nu\right) \frac{1}{\tau_{\rm UV}}.
\end{align} 
Here $\tau_{\rm UV} \gg 1$ represents the UV cutoff in our numerical computation.

Some examples of the numerical solutions of the factorized mode function $\tilde{B}(\tau)$ are given in Figure~\ref{fig.B_evolution}. One can see that $\tilde{B}$ essentially captures the oscillation dynamics led by the mass term $M_\ast = m_\ast/H$ in the late-time limit when $\tau = -k\eta \ll 1$, while the conventional vacuum mode oscillations ($\sim e^{i\tau}$) in the early-time limit ($\tau \gg 1$) has been factored out. The choice of the initial time $\tau_{\rm UV} \gg 1$ must be sufficiently large such that $\tilde{B}$ can precisely reproduce the standard analytic solution in the scale-invariant case:
\begin{align}\label{def_B0}
	\tilde{B}_0 \equiv \tilde{B}(\Delta_\epsilon = 0) = \sqrt{\tau} H_\nu^{(1)}(\tau) e^{-i\tau}.
\end{align}
Note that for the cases of $M_\ast > 3/2$, one should input $\nu = i\mu$ with $\mu = (M_\ast^2-9/4)^{1/2}$ to the initial conditions \eqref{def_B_tilde_initial_condition}. 
In Figure~\ref{fig.B_evolution}, we pick up a constant $\epsilon_U$ by taking $\Delta N = 55$ in \eqref{def_e_fold_phi} as an example so that
\begin{align}\label{def_Delta55}
	\Delta_{55} \equiv \Delta_\epsilon (\Delta N = 55) \approx 0.018.
\end{align}

There is one more advantage to consider the factorized mode function $\tilde{B}$. The late-time oscillation of the mode function led by the scalar mass $m_\ast$ is the key to generate the so-call quantum clock signals. For $\tilde{B}$, this means that the effective mass term in \eqref{eom_tilde_B} shall satisfies
\begin{align}\label{def_mass_condition}
	M_{\rm eff}^2	= \left[ \frac{M_\ast^2}{\tau^{2+\Delta_\epsilon}} \left(\frac{k}{k_\ast}\right)^{\Delta_\epsilon} - \left(1+\frac{\Delta_\epsilon}{2}\right)\left(2+\frac{\Delta_\epsilon}{2}\right)\frac{1}{\tau^2}  \right] >0,
\end{align} 
or otherwise the mass is tachyonic and the oscillation on superhorizon scales does not occur. In the standard case with $\Delta_\epsilon = 0$, this condition indicates $M_\ast^2 > 2$, where $M_\ast^2 = 2$ corresponds to conformally coupled scalars in the de Sitter space \cite{Arkani-Hamed:2015bza}. 

\begin{figure}[]
	\begin{center}
		\includegraphics[width=10 cm]{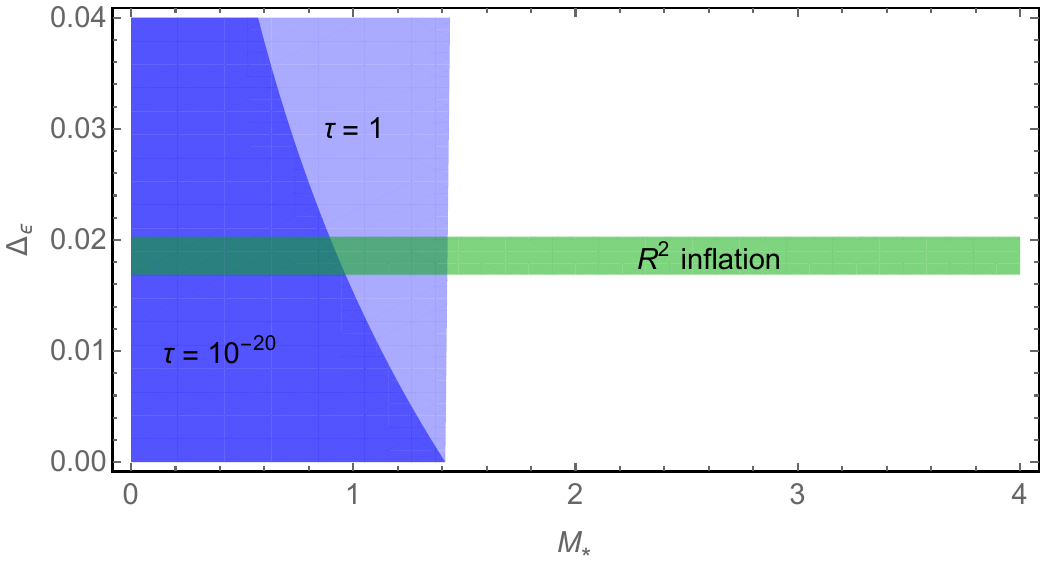}
		\hfill
		\includegraphics[width= 4.5 cm]{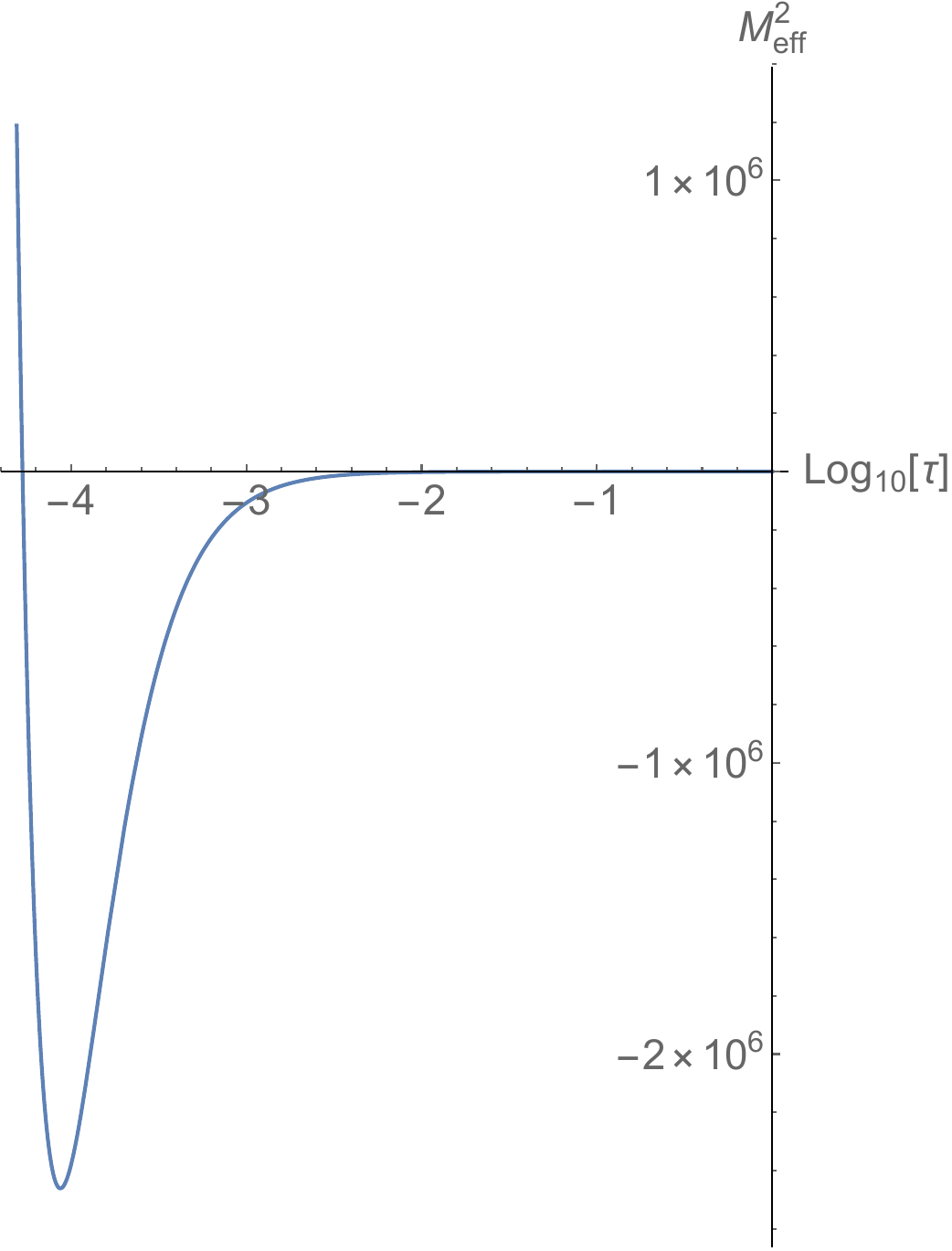}
	\end{center}
	\caption{\label{fig.mass_condition} 
		(Left Panel) The evolution of the tachyonic mass (blue-colored) region with respect to time which violates the condition \eqref{def_mass_condition}.  The green-colored window is the parameter space of constant $\Delta_\epsilon$ fixed by the value of $\epsilon_U$ in $R^2$ inflation from $\Delta N = 50$ to $\Delta N = 60$. For massive scalars with $M_\ast^2 \lesssim 2$ in $R^2$ inflation, their factorized mode functions $\tilde{B}$ do not decay after horizon crossing around $\tau = 1$ if $\Delta_\epsilon > 0$. Instead, they start to grow until very late time ($\tau \ll 1$) when \eqref{def_mass_condition} is satisfied. 
		(Right Panel) An example for the time evolution of the effective mass $M_{\rm eff}^2(\tau)$ given by \eqref{def_mass_condition} with $M_\ast = 1.3$ and $\Delta_\epsilon = \Delta_{55}$, where $M_{\rm eff}^2(1) < 0$ and $M_{\rm eff}^2(10^{-20}) >0$.
	}
\end{figure}

Interestingly, for the presence of a non-zero $\Delta_\epsilon$ the condition \eqref{def_mass_condition} becomes time dependent. An illustration of the evolution of the condition \eqref{def_mass_condition} is given in Figure~\ref{fig.mass_condition} with the window of constant $\Delta_\epsilon$ values in $R^2$ inflation ranging from the choices of $\Delta N = 50$ to $\Delta N = 60$. For the case of $\Delta_\epsilon = \Delta_{55}$, $M_\ast = 1.3$ is inside the tachyonic (blue-colored) region around the epoch of horizon crossing at $\tau = 1$. Thus the mode function $\tilde{B}$ is growing on superhorizon scales until sufficiently late time when the scalar mass $m_\ast$ finally dominates and \eqref{def_mass_condition} is satisfied. The evolution of the $M_{\rm eff}^2(\tau)$ with $M_\ast = 1.3$ is shown in the right panel of Figure~\ref{fig.mass_condition}. In this case the mode function has a peak amplitude in the region of $\tau < 1$, as shown in the left panel of Figure~\ref{fig.B_evolution}. This behavior is unfamiliar for the standard case with $\Delta_\epsilon = 0$, and it provides a possible realization of the ``cosmological tachyon collider \cite{McCulloch:2024hiz}.''
On the other hand, for $M_\ast^2 > 2$ the condition \eqref{def_mass_condition} is always satisfied for arbitrary choices of $\Delta_\epsilon$. In this case the mode function $\tilde{B}$ starts to decay after crossing the horizon around $\tau = 1$, as shown in the right panel of Figure~\ref{fig.B_evolution}. This is the usual behavior for massive scalars on superhorizon scales.


\subsection{Solutions with a slow-roll scalaron}\label{Sec_SR_scalaron}
\begin{figure}[]
	\begin{center}
		\includegraphics[width= 7.5 cm]{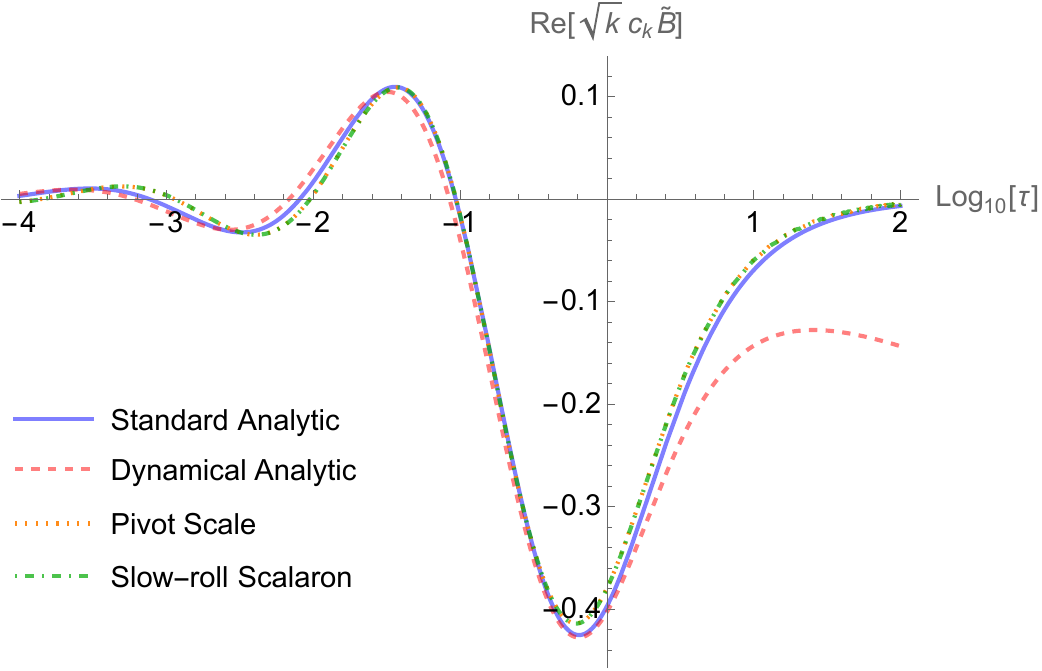}
		\hfill
		\includegraphics[width= 7.5 cm]{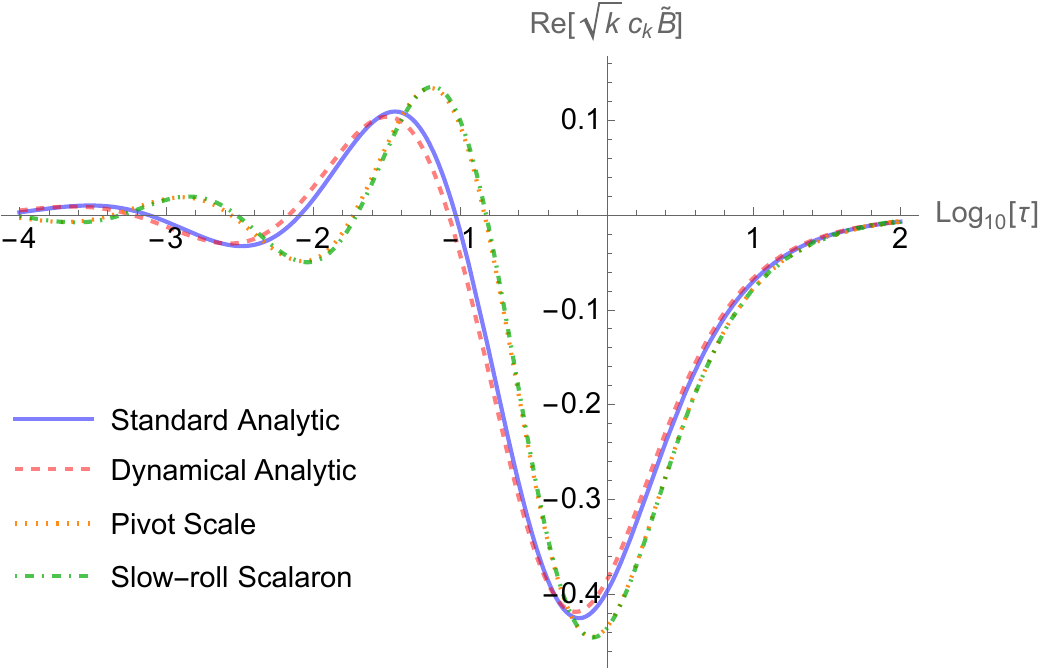}
	\end{center}
	\caption{\label{fig.B_compare} 
		Numerical solutions of the factorized mode function $\tilde{B}$ with $k/k_\ast = 1$ (Left Panel) and $k/k_\ast = 1000$ (Right Panel) based on the pivot-scale expansion (Section~\ref{Sec_constant_epsilon}) and the slow-roll scalaron approximation (Section~\ref{Sec_SR_scalaron}) with $M_\ast = 2$ and $\Delta_\epsilon = \Delta_{55}$ given by \eqref{def_Delta55}. These results are compared with analytic formulae with a standard constant mass and a dynamical time-dependent mass (Section~\ref{Sec_dynamical_mass}).
	}
\end{figure}

For a given potential in slow-roll inflationary models, we can obtain the inflaton field value at the end of inflation, $\phi_{\rm end}$, at the epoch when either one of the slow-roll parameters (defined as \ref{def_SR_parameters}) reaches to the order of unity. In $R^2$ inflation, \eqref{def_e_fold_phi} gives a good analytic approximation for the inflaton value in terms of the $e$-folding numbers to the end of inflation (see also Appendix~\ref{Append_R2}). This allows us to express the conformal coupling as
\begin{align}\label{conformal_factor_DeltaN}
	e^{-\sqrt{\frac{2}{3}} \frac{\phi_0}{M_P}} = e^{-\sqrt{\frac{2}{3}} \frac{\phi_0 -\phi_\ast + \phi_\ast}{M_P}} 
	= e^{-\sqrt{\frac{2}{3}} \frac{\phi_\ast}{M_P}} \; \frac{\Delta N_\ast}{\Delta N},
\end{align}
where $\Delta N = \ln(a_{\rm end}/a) = \ln \tau - \ln\tau_{\rm end}$ and $\tau_{\rm end} \equiv -k\eta_{\rm end}$. $\Delta N_\ast = \ln\tau_\ast - \ln\tau_{\rm end}$ is the number of $e$-folds from $k_\ast$ crosses the horizon to the end of inflation. We take $\Delta N_\ast = 55$ with $-k_\ast\eta_\ast  =1$ and $-k_\ast \eta_{\rm end} = e^{-55}$ for comparing with results of the pivot-scale expansion approach based on $\Delta_{55}$ given by \eqref{def_Delta55}. Note that $\tau_{\rm end} \ll \tau_{\rm IR}$ is required with $\tau_{\rm IR}$ being the late-time cutoff in our numerical computations.

In terms of the conformal time $\eta$, the equation of motion \eqref{eom_sigma_general} becomes
\begin{align}\label{eom_sigma_eta_SRS}
	\delta\sigma^{\prime\prime} - \frac{2+\Delta_\epsilon}{\eta} \delta\sigma^{\prime} +
	\left[k^2 + 	e^{-\sqrt{\frac{2}{3}} \frac{\phi_0}{M_P}} \frac{m_\sigma^2}{(-H\eta)^{2}}\right] \delta\sigma = 0,
\end{align}
where a prime denotes the derivative with respect to $\eta$. In terms of the expression in \eqref{conformal_factor_DeltaN} for the slow-roll inflaton as a function of the conformal time, we can import the factorized mode function \eqref{def_B_tilde} and rewrite the equation of motion following the same procedure as 
\begin{align}\label{eom_tilde_B_SR_phi}
	\frac{\partial^2}{\partial\tau^2} \tilde{B} +2i \frac{\partial}{\partial\tau} \tilde{B} +
	\left[ \frac{M_\ast^2}{\tau^{2}} \left(\frac{55}{\ln \tau - \ln x_k +55}\right) 
		- \left(1+\frac{\Delta_\epsilon}{2}\right)\left(2+\frac{\Delta_\epsilon}{2}\right)\frac{1}{\tau^2}  \right] \tilde{B} = 0,
\end{align}
where $x_k = k /k_\ast$, and $M_\ast^2 = m_\ast^2/H^2$ with $m_\ast^2 = e^{-x_\ast}m_\sigma^2$ and $x_\ast = \sqrt{2/3} \;\phi_\ast/M_P$ is the same definition as used in \eqref{eom_tilde_B}. The solution depends on, but is not very sensitive to the choice of $50 \leq \Delta N_\ast \leq 60$. 
Initial conditions of $\tilde{B}$ shares the same as used in \eqref{def_B_tilde_initial_condition}.

\subsection{Analytic formulae with a dynamical mass}\label{Sec_dynamical_mass}
As a reference, we examine the analytic formalism for the scalar mode functions with a time-dependent mass provided in \cite{Aoki:2020wzu}. In this approach, we simply start with a general assumption for the slow-roll inflation as $\dot{\phi}_0 = (-H\eta)\phi_0^{\prime} = \sqrt{2\epsilon_1}M_PH$. Taking $\epsilon_1$ and $H$ as constants, we can work out $\phi_0$ as a function of the conformal time $\eta$, where $\phi_0(\eta) = \sqrt{2\epsilon_1}M_P\ln(\eta/\eta_0)$ for some initial epoch $\eta_0$. Expanding the function near a pivot scale $k_\ast = -1/\eta_\ast$, we get
\begin{align}
	\phi_0(\eta) = \phi_\ast -\sqrt{2\epsilon_1} M_P (1- \frac{\eta}{\eta_\ast}) + \cdots,
\end{align}  
where $\phi_\ast = \phi_0(\eta_\ast)$. Applying this expansion to the conformal coupling with mass, we can obtain
\begin{align}
	e^{-\alpha \phi_0/M_P} \frac{m^2}{H^2} = M_\ast^2 \left(1-\alpha\sqrt{2\epsilon_1}\right) + M_\ast^2 \alpha\sqrt{2\epsilon_1} \frac{\eta}{\eta_\ast} + \cdots,
\end{align}
where $M_\ast^2 = m^2/H^2$ is aligned with solutions in previous methods.

For the discussion of $R^2$ inflation, we impose $\alpha = \sqrt{2/3}$ and $\Delta_\epsilon = \alpha \sqrt{2\epsilon_1}$. 
The equation of motion for the perturbations of a scalar $\sigma$ with a dynamical mass reads
\begin{align}\label{eom_sigma_eta_dM}
	\delta\sigma^{\prime\prime} - \frac{2+\Delta_\epsilon}{\eta} \delta\sigma^{\prime} +
	\left[k^2 + 	 \frac{M_\ast^2}{\eta^{2}}\left(1-\Delta_\epsilon\right) + \Delta_\epsilon \frac{M_\ast^2}{\eta\eta_\ast}  \right] \delta\sigma = 0.
\end{align}
Again, it is convenient to do the transfer $\delta\sigma = (-k\eta)^{1+\Delta_\epsilon/2}\delta\tilde{\sigma}$ and rewrite the equation of motion as
\begin{align}
	\frac{\partial^2}{\partial\tau^2} \delta\tilde{\sigma} + \left[1 + \frac{A_1}{\tau^2} + \frac{A_2}{\tau}\right] \delta\tilde{\sigma} = 0,
\end{align}
where we have defined the parameters
\begin{align}
	A_1 \equiv M_\ast^2(1-\Delta_\epsilon) - \left(1+\frac{\Delta_\epsilon}{2}\right) \left(2+\frac{\Delta_\epsilon}{2}\right), 
	\quad A_2 \equiv \Delta_\epsilon M_\ast^2 \frac{k_\ast}{k}.
\end{align}
This equation can be solved analytically with respect to the Bunch-Davies vacuum. The mode function is given by \cite{Aoki:2020wzu}:
\begin{align}\label{tilde_u_dM}
	\tilde{u}_k = -\frac{i}{\sqrt{2k}} e^{\frac{\pi}{4}A_2} \left(2\tau\right)^{iA_2/2} W_{-iA_2/2, i \sqrt{A_1-1/4}} (-2i\tau),
\end{align}
where $W_{a,b}(Z)$ is the Whittaker W function. 

The comparison of numerical solutions from the pivot-scale expansion (Section~\ref{Sec_constant_epsilon}) and the slow-roll inflaton approximation (Section~\ref{Sec_SR_scalaron}) with analytic mode functions based on the standard constant mass and time-dependent dynamical mass (this section) is given in Figure~\ref{fig.B_compare}. The standard analytic solution is protected by the dilatation symmetry so that it is invariant under the choice of $k$. For the choice of $k/k_\ast \gg 1$, we find that both numerical solutions agree with each other. 

\section{Corrections to the power spectrum}\label{Sec_Power_Spectrum}

In this section we examine the corrections to the power spectrum $P_{\delta\phi}$ of $\phi$ led by the scalar perturbation $\delta\sigma$ through the transfer vertex \eqref{def_L2}. This examination can provides a consistency check with the observational constraints  \cite{Arkani-Hamed:2018kmz}. In the limit of $\Delta_\epsilon = 0$ for the near pivot-scale expansion method (Section~\ref{Sec_constant_epsilon}), we can check if the result is also consistent with the closed-form formulae of quasi-single-field inflation presented in \cite{Chen:2012ge}.

 The power spectrum can be obtained from the expectation values of $\delta\phi^2$ at the end of inflation through $\langle \delta\phi_{\vec{k}_1}\delta\phi_{\vec{k}_2}\rangle = (2\pi)^3\delta^{(3)}(\vec{k}_1 + \vec{k}_2) (2\pi^2/k_1^3) P_{\delta\phi}$, where we adopt the standard analytic form for a massless inflaton mode function as
 \begin{align}\label{def_mode_inflaton}
 	\delta\phi_{\vec{k}}(\eta) \equiv f_k(\eta) \hat{a}_{\vec{k}} +f_k^\ast(\eta) \hat{a}_{-\vec{k}}^\dagger, \qquad f_k(\eta) =\frac{H}{\sqrt{2k^3}} \left(1+ik\eta\right) e^{-ik\eta}.
 \end{align}
 On the other hand, the mode functions of the massive scalar are solved numerically as
 \begin{align}
 	\delta\sigma_{\vec{k}}(\eta) \equiv u_k(\eta) \hat{b}_{\vec{k}} + u_k^\ast(\eta) \hat{b}_{-\vec{k}}^\dagger
 	 = \frac{1}{a^{1+\Delta_\epsilon/2}} \left[ \tilde{u}_k(\eta) \hat{b}_{\vec{k}} + \tilde{u}_k^\ast(\eta) \hat{b}_{-\vec{k}}^\dagger\right],
 \end{align}
 where $\tilde{u}_k = c_k \tilde{B} e^{-ik\eta}$ and $\tilde{B}(k,\eta)$ is solved by \eqref{eom_tilde_B} or \eqref{eom_tilde_B_SR_phi}. 
 
 To compute the correlation functions based on the  in-in formalism \cite{Weinberg:2005vy,Chen:2010xka}, we shall construct the interaction picture of our system. 
 As a first step, we need to identify the Hamiltonian density $\mathcal{H}[\delta\phi, \delta\sigma, \delta\pi_\phi, \delta\pi_\sigma]$ in terms of the conjugate momentum densities $\delta\pi_\phi = \partial\delta\mathcal{L}/\partial\delta\dot{\phi}$, $\delta\pi_\sigma = \partial\delta\mathcal{L}/\partial\delta\dot{\sigma}$, where $\delta\mathcal{L}$ describes the perturbed Lagrangian. 
 At quadratic order of the linear perturbations, the free-field Lagrangian reads
 \begin{align}
 	\delta\mathcal{L}_2 =& \frac{a^3}{2} \left[\left(\delta\dot{\phi}\right)^2 - \frac{1}{a^2} \left(\partial_i\delta\phi\right)^2\right] \\\nonumber
&+  \frac{a^3}{2} e^{-\sqrt{\frac{2}{3}} \frac{\phi_0}{M_P}}\left[\left(\delta\dot{\sigma}\right)^2 - \frac{1}{a^2} \left(\partial_i\delta\sigma\right)^2\right]
-\frac{a^3}{2} e^{-2\sqrt{\frac{2}{3}} \frac{\phi_0}{M_P}} m_\sigma^2 \delta\sigma^2.
 \end{align} 
 This gives the conjugate momentum densities in the interaction picture as
 \begin{align}
 	\delta\pi_\phi = a^3\delta\dot{\phi}, \qquad \delta\pi_\sigma = a^3  e^{-\sqrt{\frac{2}{3}} \frac{\phi_0}{M_P}} \delta\dot{\sigma}.
 \end{align}
 After separating the Hamiltonian density into the kinematic part $\mathcal{H}_0$ and the interaction part $\mathcal{H}_I$, we replace the perturbations by the ones defined in the interaction picture, namely $\delta\phi_I$ and $\delta\sigma_I$, as solutions of the equation of motion given by $\mathcal{H}_0$.
 As a final step, we use $\delta\dot{\phi}_I = \partial\mathcal{H}_0/\partial\delta\pi_\phi$ and $\delta\dot{\sigma}_I = \partial\mathcal{H}_0/\partial\delta\pi_\sigma$ to replace the momentum densities so that the Hamiltonian densities are given by
 \begin{align}
 	\mathcal{H}_0 =& \frac{a^3}{2} \left[\left(\delta\dot{\phi}_I\right)^2 + \frac{1}{a^2} \left(\partial_i\delta\phi_I\right)^2\right] \\\nonumber
 	&+  \frac{a^3}{2} e^{-\sqrt{\frac{2}{3}} \frac{\phi_0}{M_P}}\left[\left(\delta\dot{\sigma}_I\right)^2 + \frac{1}{a^2} \left(\partial_i\delta\sigma_I\right)^2\right]
 	+\frac{a^3}{2} e^{-2\sqrt{\frac{2}{3}} \frac{\phi_0}{M_P}} m_\sigma^2 \delta\sigma_I^2, 
 	\\\label{def_H2}
 	\mathcal{H}_{I2} =& -\frac{c_2}{M_P^2} a^3 e^{-\sqrt{\frac{2}{3}} \frac{\phi_0}{M_P}} \dot{\phi}_0\sigma_0 \delta\dot{\phi}_I \delta\sigma_I.
 \end{align}
We can replace the time-dependent conformal coupling by $e^{-\sqrt{2/3}\phi_0/M_P} = e^{-x_\ast} (a/a_\ast)^{\Delta_\epsilon} $ for the near pivot-scale expansion approach and by $e^{-\sqrt{2/3}\phi_0/M_P} = e^{-x_\ast} (\Delta N_\ast/\Delta N)$ for the slow-roll scalaron approximation, where $x_\ast = \sqrt{2/3}\;\phi_\ast/M_P$.
 
To compute the $n$-point correlation function by virtue of the in-in formalism, we perform the expansion:
\begin{align}
\left\langle \delta\phi^n\right\rangle^\prime &=
\langle 0 \vert \left[\bar{T}\exp\left(i\int_{t_0}^{t}dt^\prime H_I(t^\prime)\right)\right]
\delta\phi_I^n(t) 
\left[T \exp\left(-i \int_{t_0}^{t} dt^{\prime\prime} H_I(t^{\prime\prime})\right)\right] \vert 0\rangle^\prime \nonumber
\\\label{2_point_non_timeordered}
&= \left\langle 0\vert\delta\phi_I^n \vert 0 \right\rangle^\prime + \int_{t_0}^{t}d\tilde{t}_1\int_{t_0}^{t} dt_1 \langle 0 \vert H_I(\tilde{t}_1)\delta\phi_I^n H_I(t_1) \vert 0 \rangle^\prime 
\\\label{2_point_timeordered}
&\qquad -2 \Re \left[\int_{t_0}^{t}dt_1\int_{t_0}^{t_1}dt_2 \langle 0\vert \delta\phi_I^n H_I(t_1)H_I(t_2)\rangle^\prime \right] + \cdots,
\end{align}
where $\langle\cdots\rangle^\prime = (2\pi)^3\delta(\vec{k}_1 + \vec{k}_2 +\cdots)\langle\cdots\rangle$, and the leading corrections are given by the ``non-time-ordered'' term  \eqref{2_point_non_timeordered} and the ``time-ordered'' term \eqref{2_point_timeordered}.

Taking $n = 2$ with $H_I = \int d^3\vec{x}\, \mathcal{H}_{I2}$, the correction to the two-point function based on the pivot-scale (PS) expansion approach reads
\begin{align}
	\left\langle \delta\phi^2\right\rangle^\prime &-  \left\langle 0\vert\delta\phi_I^2 \vert 0 \right\rangle^\prime  =
	c_2^2\; e^{-2x_\ast}\left(\frac{k_\ast}{H}\right)^{-2\Delta_\epsilon} \frac{\sigma_0^2\dot{\phi}_0^2}{M_P^4} \frac{H^2}{2k^3} 
	\left\{\left\vert \int_{-\infty}^{0} d\eta\, a(\eta)^{3+\Delta_\epsilon}f_k^\prime(\eta) u_k(\eta)\right\vert^2 \right. \\\nonumber
	 &\qquad \left.  -2\Re \left[\int_{-\infty}^{0}d\eta_1 a(\eta_1)^{3+\Delta_\epsilon}f_k^{\ast\prime}(\eta_1) u_k(\eta_1) 
	 \int_{-\infty}^{\eta_1} d\eta_2 a(\eta_2)^{3+\Delta_\epsilon}f_k^{\ast\prime}(\eta_2) u_k^\ast(\eta_2) \right] \right\} \\
	 \label{2pt_final}
	 &= c_2^2 \;e^{-2x_\ast}\left(\frac{k_\ast}{H}\right)^{-2\Delta_\epsilon} \frac{\epsilon_U}{2} \frac{\sigma_0^2}{M_P^2} 
	 \left(\frac{k}{H}\right)^{2\Delta_\epsilon} \frac{H^2}{k^3}\mathcal{C}_{\rm PS}(\Delta_\epsilon, M_\ast, x_k) , 
	 \nonumber\\
	 &= c_2^2 \;e^{-2x_\ast} \epsilon_U \frac{\sigma_0^2}{M_P^2}\;\frac{H^2}{2k^3} C(\Delta_\epsilon, M_\ast, x_k)
\end{align}
where we have used $\epsilon_U\approx \dot{\phi}^2_0/(2M_P^2H^2)$ in \eqref{2pt_final} and the numerical factor $C(\Delta_\epsilon, M_\ast, x_k) = x_k^{2\Delta_\epsilon} \mathcal{C}_{\rm PS}(\Delta_\epsilon, M_\ast, x_k)$ with $x_k \equiv k/k_\ast$. In the standard case with $\Delta_\epsilon = 0$ the factor $\mathcal{C}_{\rm PS}$ coincides with $C$. 
Taking $c_2 = 1$ and using $\Delta_\epsilon = \Delta_{55}$, $\epsilon_U = 2.6 \times 10^{-4}$, $\phi_\ast/M_P = 5.26$ based on $\Delta N = 55$ in $R^2$ inflation, we find
\begin{align}
	\left\langle \delta\phi^2\right\rangle^\prime -  \left\langle 0\vert\delta\phi_I^2 \vert 0 \right\rangle^\prime  
	\sim  10^{-8}  \frac{\sigma_0^2}{M_P^2}\; \frac{H^2}{2k^3} \times C(\Delta_{55}, M_\ast, 1).
\end{align}
Here $\left\langle 0\vert\delta\phi_I^2 \vert 0 \right\rangle^\prime = H^2/2k^3$ is the zeroth-order expectation value of the two-point function.
The VEV of $\sigma_0$ depends on the model of the isocurvature scalar, yet, in general we expect $\sigma_0/M_P \ll 1$. 

Let us focus on the numerical factor $C = x_k^{2\Delta_\epsilon} \mathcal{C}_{\rm PS}$ in \eqref{2pt_final}, which collects the contribution that is independent of the coupling constants of the transfer vertices. 
In the pivot-scale expansion approach, the factor $\mathcal{C}_{\rm}(\Delta_\epsilon, M_\ast, x_k)$ is defined as
\begin{align}\label{def_C}
 \mathcal{C}_{\rm PS}	\equiv  \frac{k^3}{H^2}  \Re 
 \left[\int_{0}^{\infty} d\tau_1 \tau_1^{-2-\Delta_\epsilon} u_k(\tau_1) \left(e^{i\tau_1} - e^{-i\tau_1}\right) \int_{\tau_1}^{\infty}d\tau_2 \tau_2^{-2-\Delta_\epsilon} u_k^\ast(\tau_2) e^{-i\tau_2} \right], 
\end{align}
where such a definition can reproduce the standard analytic results found in \cite{Chen:2012ge} when employing the mode function for a massive scalar with
\begin{align}\label{def_standard_u_mode}
	u_{k} = c_k a^{-1} \sqrt{\tau} H_{i\mu}^{(1)}(\tau) = c_k\frac{H}{k} \tau^{3/2} H_{i\mu}^{(1)}(\tau). 
\end{align}
Note that \eqref{def_C} is a combined representation for both the time-ordered term \eqref{2_point_timeordered} and the non-time-ordered term \eqref{2_point_non_timeordered}. One can use the trick $\int_{0}^{\infty}dx_1\int_{0}^{\infty}dx_2 = \int_{0}^{\infty}dx_1 (\int_{x_1}^{\infty} + \int_{0}^{x_1})dx_2$ to the non-time-ordered term and change the order of integration to obtain this represetation. Such a represtation is also used in \cite{Chen:2009zp} for the standard $\Delta_\epsilon = 0$ case with $\mu = -i\nu$. It is clear to see that the factor $e^{i\tau} - e^{-i\tau}$ in the outer integral of \eqref{def_C} can cancel out the divergence in the late-time limit with $\tau \rightarrow 0$. 

\begin{figure}[]
	\begin{center}
		\includegraphics[width=12 cm]{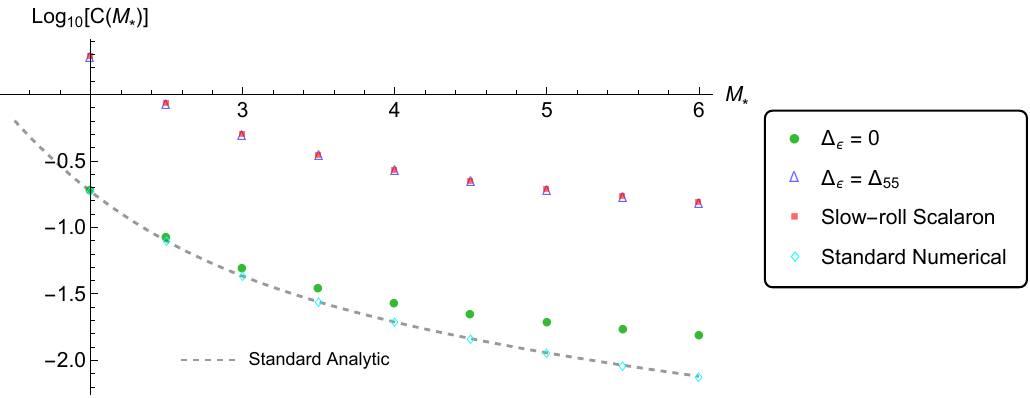}
	\end{center}
	\caption{\label{fig.C} 
		The numerical factor $C(M_\ast)$ for the correction to the inflaton power spectrum with $k = k_\ast$. 
		Results of the near pivot-scale expansion are presented with $\Delta_\epsilon = 0$ for the standard dilatation invariant case and $\Delta_\epsilon = \Delta_{55}$ for the $R^2$ inflation. The scale dependence in the modes functions solved with the slow-roll scalaron cannot be turned off. For a consistency check with the analytic formula \eqref{def_C_analytic}, we provide the numerical results for the standard analytic mode functions using the same set off cutoffs $\{\tau_{\rm IR}, \tau_{\rm UV}\}$ for all the numerical computations. 
	}
\end{figure}

In terms of the factorized mode function $\tilde{B}$, the numerical factor is given by
\begin{align}\label{def_C_B_tilde}
	C(\Delta_\epsilon, & M_\ast, x_k)	= x_k^{2\Delta_\epsilon} \mathcal{C}_{\rm PS}(\Delta_\epsilon, M_\ast, x_k) \\\nonumber
	&= \frac{\pi}{4}e^{-\pi \mu} x_k^{\Delta_\epsilon}\left( \frac{k_\ast}{H}\right)^{-\Delta_\epsilon} \\\nonumber
	&\;\;\times 
	\Re 
	\left[\int_{0}^{\infty} d\tau_1 \tau_1^{-1-\Delta_\epsilon/2} \tilde{B}(x_k, \tau_1) \left(e^{2i\tau_1} - 1\right) \int_{\tau_1}^{\infty}d\tau_2 \tau_2^{-1-\Delta_\epsilon/2} \tilde{B}^\ast(x_k, \tau_2) e^{-2i\tau_2} \right], 
\end{align}
for the solution of \eqref{def_B_tilde}. 

On the other hand, for the slow-roll (SR) scalaron approximation we can follow the same procedure to obtain
\begin{align}
\left\langle \delta\phi^2\right\rangle^\prime -  \left\langle 0\vert\delta\phi_I^2 \vert 0 \right\rangle^\prime  
= c_2^2 \;e^{-2x_\ast} \epsilon_U \frac{\sigma_0^2}{M_P^2}\;\frac{H^2}{2k^3} C(\Delta_\epsilon, M_\ast, x_k),
\end{align}
with the numerical factor defined as
\begin{align}\label{def_C_B_tilde_SRS}
	C(\Delta_\epsilon, & M_\ast, x_k) = \mathcal{C}_{\rm SR} 
	= \frac{\pi}{4}e^{-\pi \mu} x_k^{-\Delta_\epsilon}\left( \frac{k_\ast}{H}\right)^{-\Delta_\epsilon} \\\nonumber
	&\times 
	\Re 
	\left[\int_{0}^{\infty} d\tau_1\; \tau_1^{-1+\Delta_\epsilon/2} \left(\frac{55}{\ln\tau_1 -\ln x_k + 55}\right) \tilde{B}(x_k, \tau_1) \left(e^{2i\tau_1} - 1\right) \right.
	\\\nonumber
	&\quad\times 
	\left. \int_{\tau_1}^{\infty}d\tau_2\; \tau_2^{-1+\Delta_\epsilon/2}  \left(\frac{55}{\ln\tau_2 -\ln x_k + 55}\right) \tilde{B}^\ast(x_k, \tau_2) e^{-2i\tau_2} \right], 
\end{align}
The results of $C(M_\ast)$ computed from different approaches are given in Figure~\ref{fig.C}.

The standard analytic result based on the dilation invariant mode function \eqref{def_standard_u_mode} is found to be of the form \cite{Chen:2012ge}:
\begin{align}\label{def_C_analytic}
		\mathcal{C}_0(M_\ast) = C(M_\ast)	=& \frac{\pi^2}{4 \cosh^2(\pi \mu)} + 
		\frac{e^{\pi \mu}}{16 \sinh(\pi \mu)} \Re\left[\psi^{(1)}\left(\frac{3}{4}+i\frac{\mu}{2}\right) -\psi^{(1)}\left(\frac{1}{4}+i\frac{\mu}{2}\right)  \right] 
		\nonumber\\
		&-\frac{e^{-\pi \mu}}{16 \sinh(\pi \mu)}   \Re\left[\psi^{(1)}\left(\frac{3}{4}-i\frac{\mu}{2}\right) -\psi^{(1)}\left(\frac{1}{4}-i\frac{\mu}{2}\right)  \right],
\end{align}
where $\psi^{(1)}(z) = d^2\ln\Gamma(z)/dz^2$ is the first derivative of the digamma function.

In Figure~\ref{fig.C}, we compare numerical results of $C(\Delta_\epsilon, M_\ast, x_k)$ from both methods of the near pivot-scale expansion and the slow-roll scalaron approximation. Their results in $R^2$ inflation agree with each other. To check the numerical error led by the early-time (or the late-time) cutoff $\tau_{\rm UV}$ ($\tau_{\rm IR}$), we perform the numerical check of \eqref{def_C_B_tilde} with $\tilde{B}$ replaced by  the standard analytic mode functions \eqref{def_B0} but using the same set of $\{\tau_{\rm IR},\tau_{\rm UV}\}$. Since the solution of $\Delta_\epsilon = 0$ is supposed to reproduce the standard analytic result, one can see that the error enhances with the increase of $M_\ast$ and therefore we shall truncate the analysis around $M_\ast \lesssim 6$. 
This numerical error is mainly due to the fast oscillations near the cutoff $\tau_{\rm UV}$, which is independent of our choices of $\Delta_\epsilon$ or $x_k$.
There is an universal enhancement $C(\Delta_{55}, M_\ast, x_k)/C(0, M_\ast, x_k) \sim (k_\ast/H)^{-\Delta_{55}}$ led by the modified decay rate of the scalar perburbation $\delta\sigma$ with a non-zero $\Delta_\epsilon$. The maximal enhancement is given by
\begin{align}\label{enhancement_factor}
	\left(\frac{k_\ast}{H}\right)^{-\Delta_{55}} \approx 
	10.1 \times \left(\frac{k_\ast}{0.002\; \textrm{Mpc}^{-1}} \frac{3\times10^{13}\;\textrm{GeV}}{H}\right)^{-\Delta_{55}},
\end{align}
where $H < 3.04 \times 10^{13}$ GeV, or the tensor-to-scalar ratio $r < 0.036$ is a recently improved upper bound \cite{BICEP:2021xfz}.
One can check the scale-dependence in terms of $x_k = k/k_\ast$ for $C(\Delta_\epsilon, M_\ast, x_k)$ is red-tilted and is the same for the two numerical approaches.

\section{The simplest quantum primordial clock}\label{Sec_Bispectrum}

Bispectrum from the 3-point correlation function of the curvature perturbation is the simplest observable for the oscillatory momentum scaling features created by the (virtual) production of a massive particle \cite{Arkani-Hamed:2015bza}. These oscillatory features are imprints of the quantum interference between vacuum fluctuations of the inflaton ($\sim e^{-ik\eta}$) and the late-time oscillation of the massive particle mode functions on superhorizon scales ($\sim e^{i m t}$). In the ideal cosmological collider with exact dilatation invariance, such an oscillatory feature is used as an ``standard clock'' to test the time evolution of the scale factor $a(t)$ in different scenarios for the primordial Universe  \cite{Chen:2015lza}.
 In this work we only focus on the scenario of inflation, $a(t) = e^{Ht}$, for the primordial Universe, but we compare the simplest quantum clock signals in $R^2$ inflation to the prediction in the ideal cosmological collider built with exact dilatation invariance. By ``simplest,'' we mean:
\begin{list}{a}{}
	\item [(1)] It resides in the primordial bispectrum,
	\item[(2)] the generation of such a signal uses the minimal number of interaction vertices, and
	\item[(3)] the shape function of this clock signal can be constructed out of one unified time integration formula.
	\footnote{This in fact relies on taking the time derivative of the inflaton perturbation in the two-point transfer vertex \eqref{def_H2} while picking up the lowest-order (or the zeroth-order) terms with respect to the differential boundary operator \eqref{def_boundary_operator} in the three-point interaction \eqref{def_H3} as if $\delta\phi$ is replaced by a conformally coupled scalar (c.f. Section 4.1 in \cite{Arkani-Hamed:2015bza}).}
\end{list}
As given by the diagrammatic illustration in the right panel of Figure~\ref{fig.Feynman_clock}, the simplest quantum primordial clock is realized with only two vertices. The two-point transfer vertex given by \eqref{def_H2} is used in the computation of the power spectrum, and the three-point vertex comes from \eqref{def_L3} reads
\begin{align}\label{def_H3}
	\mathcal{H}_{I3} = - \frac{c_3}{M_P^2} e^{-x_0} a^{3} \sigma_0
	\left[\left(\delta\dot{\phi}_I\right)^2 +\frac{1}{a^2}\left(\partial_i\delta\phi_I\right)^2\right] \delta\sigma_I,
\end{align}
where $x_0 = \sqrt{\frac{2}{3}}\frac{\phi_0}{M_P}$ and the time-dependence of $x_0$ (or $\phi_0$) in the two numerical approaches are given
\begin{align}
	e^{-x_0} = e^{-x_\ast}\left(\frac{a}{a_\ast}\right)^{\Delta_\epsilon} = e^{-x_\ast} \left(\frac{\Delta N_\ast}{\Delta N}\right),
\end{align}
respectively.
Here $a_\ast = k_\ast/H$ and $\Delta N_\ast = \ln(a_{\rm end}/a_\ast) = \ln(k_{\rm end}/k_\ast) = 55$ is used.

Following the in-in formalism, the (simplest) quantum clock signal can be obtained via 
\begin{align}
	\left\langle \delta\phi^3\right\rangle^\prime =&
	 \int_{t_0}^{t}d\tilde{t}_1\int_{t_0}^{t} dt_1 \langle 0 \vert H_I(\tilde{t}_1)\delta\phi_I^3 H_I(t_1) \vert 0 \rangle^\prime 
	\label{3pt_nontime_ordered} \\ \label{3pt_time_ordered}
	 &\qquad -2 \Re \left[\int_{t_0}^{t}dt_1\int_{t_0}^{t_1}dt_2 \langle 0\vert \delta\phi_I^3 H_I(t_1)H_I(t_2) \vert 0 \rangle^\prime \right] +\cdots,
\end{align}
assuming that $\delta\phi_I$ is a Gaussain random field. 

In the computation of this section, we assign $k_3$ as the external momentum leg fixed by the transfer vertex so that it is conserved with the internal momentum of the massive scalar perturbation where $\delta\sigma_{k} = \delta\sigma_{k_3}$. We always align the equilateral configuration of the kinematic triangle with respect to the pivot scale, such that $k_1 = k_2 = k_3 = k_\ast$. It is possible to obtain the squeezed configuration from (a) $k_1 = k_2 = k_\ast$ with $k_3 \ll k_\ast$ or (b) $k_1 = k_2 \gg k_\ast$ with $k_3 = k_\ast$. When the dilatation invariance is broken by the exchange process with the massive scalar perturbation $\delta\sigma$ in $R^2$ inflation, the results from (a) and (b) can be different. We refer the squeezed configuration of (a) as the ``forward'' scaling since it measures the scales exit the horizon earlier than $k_\ast$, and thus the configuration of (b) is referred as the ``backward'' scaling.
Note that squeezing the external leg for $k_1$ or $k_2$ does not lead to important ``clock'' signals since the quantum interference only occurs with the mass-induced oscillations in the scalar mode functions $u_k$ on scales well outside the horizon. 


\subsection{Shape functions with analytic calibrations}

\begin{figure}[]
	\begin{center}
		\includegraphics[width= 14 cm]{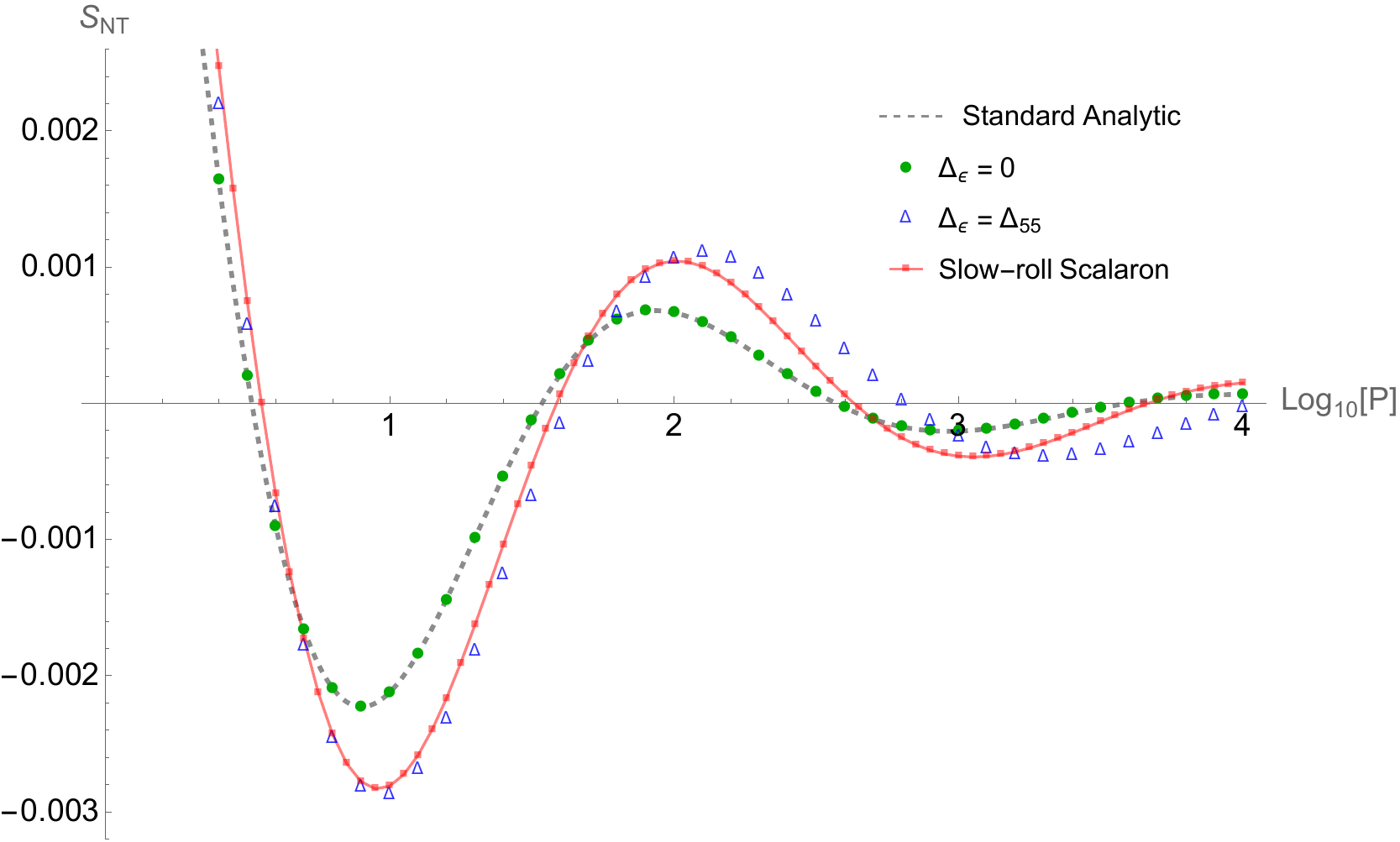}
		\hfill
		\includegraphics[width= 14 cm]{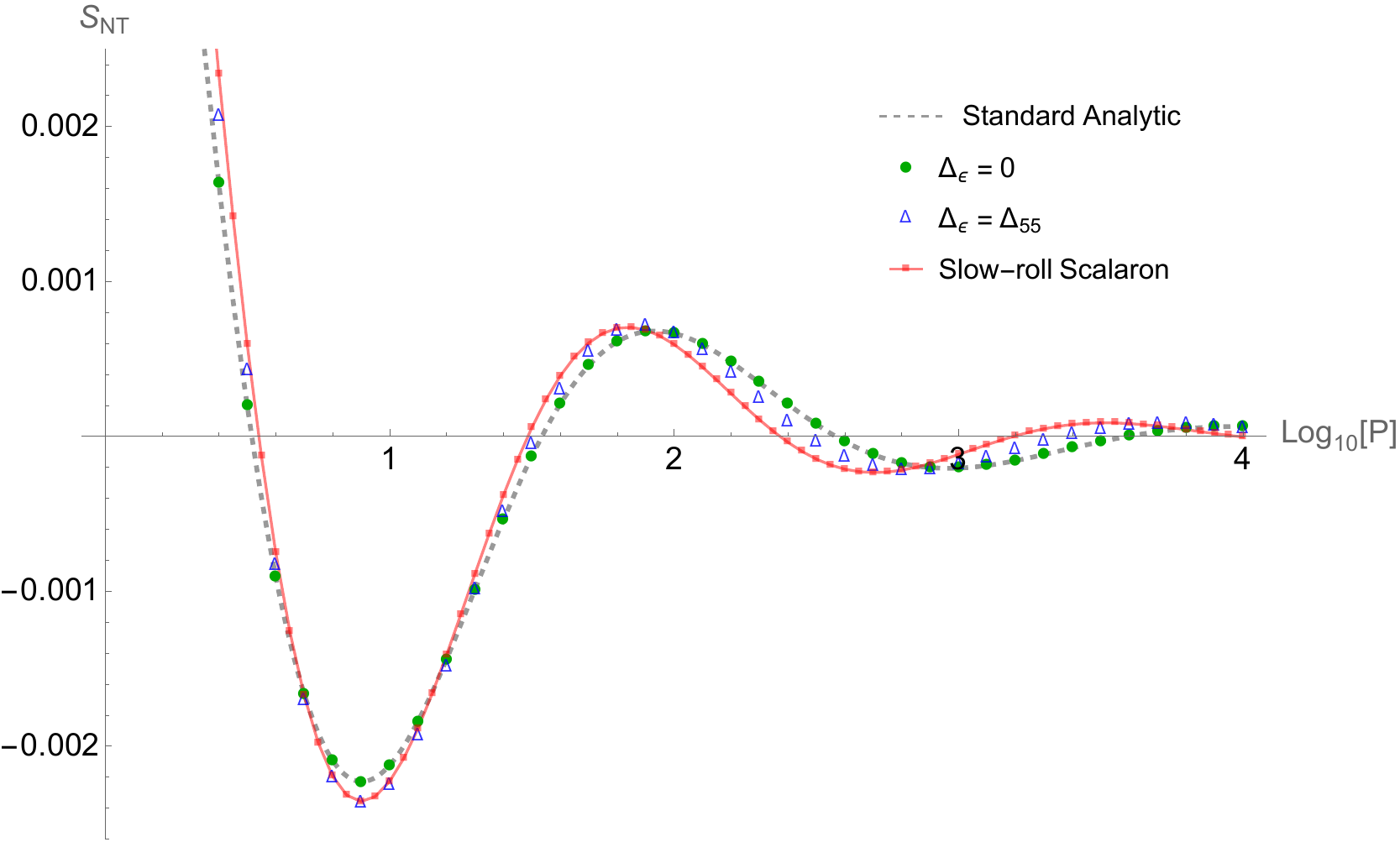}
	\end{center}
	\caption{\label{fig.S_non_time_ordered} 
		The model-independent shape function from the non-time-ordered contribution \eqref{3pt_nontime_ordered} for (Upper Panel) the forward-scaling triangle configuration with $k_1 = k_2 = k_\ast$, $k_3 \leq k_\ast$ and (Lower Panel) the backward-scaling triangle configuration with $k_1 = k_2 \geq k_\ast$, $k_3 = k_\ast$. Results from the pivot-scale expansion approach $S_{\rm PS}$ are given with the choices of $\Delta_\epsilon = \Delta_{55}$ and $\Delta_\epsilon = 0$. The results of $\Delta_\epsilon = 0$ in the two panels are in well agreement with the standard analytic prediction and theferore they are identical in the two scaling configurations. 
		We use $M_\ast = 2$ in this plot.
	}
\end{figure}

Let us focus on the non-time-ordered term \eqref{3pt_nontime_ordered}, which is found to be a subdominant contribution to the ``quantum primordial standard clock \cite{Chen:2015lza}'' in the large mass limit $m/H \gg 1$ when comparing to the contribution from the time-ordered part \eqref{3pt_time_ordered}. However, as we will see, the spectral shape generated by the non-time-ordered term is a pure oscillatory function, while the shape function result from the time-ordered integration is in general dominated by some non-oscillatory components. Therefore, the non-time-ordered results can provide a better visualization of the ``clock signal.''
Furthermore, in the standard case with exact symmetries, the non-time-ordered integration can be worked out analytically. This allows us to check if our numerical computation can recover a consistent result by turning off the parameter $\Delta_\epsilon$. For the purpose of calibration with the standard analytic formalism, we can only adopt the near pivot-scale expansion approach given by Section~\ref{Sec_constant_epsilon} since the explicit $k$-dependence in the mass term of the slow-roll scalaron approximation in Section~\ref{Sec_SR_scalaron} cannot be switched off. 

The non-time-ordered term \eqref{3pt_nontime_ordered} is composed by the two oppisite time orderings for the interaction vertices given in Figure~\ref{fig.Feynman_clock}:
\begin{align}
	\int_{t_0}^{t}d\tilde{t}_1\int_{t_0}^{t} dt_1 &\langle 0 \vert H_I(\tilde{t}_1)\delta\phi_I^3 H_I(t_1) \vert 0 \rangle^\prime =\\\nonumber
	&\int_{t_0}^{t}d\tilde{t}_1\int_{t_0}^{t} dt_1 
	\left[ \langle 0 \vert H_{I2}(\tilde{t}_1)\delta\phi_I^3 H_{I3}(t_1) \vert 0 \rangle^\prime +\langle 0 \vert H_{I3}(\tilde{t}_1)\delta\phi_I^3 H_{I2}(t_1) \vert 0 \rangle^\prime \right],
\end{align}
where these two terms are complex conjugate to each other. To see this, let us first work out the two-point anti-time-ordering and the three-point time-ordering correlation for the pivot-scale expansion approach as
\begin{align}
\int_{t_0}^{t}  & d\tilde{t}_1\int_{t_0}^{t} dt_1   \langle 0 \vert H_{I2}(\tilde{t}_1)\delta\phi_I^3 H_{I3}(t_1) \vert 0 \rangle^\prime = \\\nonumber
& c_2c_3 \; e^{-2x_\ast} \left(\frac{k_\ast}{H}\right)^{-2\Delta_\epsilon}  \frac{\sigma_0^2\dot{\phi}_0}{M_P^4} 
f_{k_1}(0)f_{k_2}(0)f_{k_3}^\ast(0) 
\int_{-\infty}^0d\tilde{\eta}_1 a^{3+\Delta_\epsilon}(\tilde{\eta}_1)f_{k_3}^\prime(\tilde{\eta}_1)u_{k_3}(\tilde{\eta}_1) 
\\\nonumber
&\qquad \times \int_{-\infty}^0 d\eta_1 a^{2+\Delta_\epsilon}(\eta_1) 
\left[f_{k_1}^{\ast\prime}(\eta_1) f_{k_2}^{\ast\prime}(\eta_1) + \vec{k}_1\cdot\vec{k}_2 f_{k_1}^\ast(\eta_1) f_{k_2}^\ast(\eta_1) \right] u_{k_3}^\ast(\eta_1), 
\end{align}
with $f_k(0) = H/\sqrt{2k^3}$ being the expectation value of \eqref{def_mode_inflaton} at the end of inflation.  
Thanks to the conformal structure preserved in the inflaton sector (as a massless scalar), it is possible to replace the time evolution of the mode function $f_k$ in the bulk by the differential operators defined on the late-time boundary surface according to \cite{Arkani-Hamed:2015bza}:
\begin{align}
	f_{k_1}^{\ast\prime}(\eta_1) f_{k_2}^{\ast\prime}(\eta) 
	&+ \vec{k}_1\cdot\vec{k}_2 \;f_{k_1}^\ast(\eta_1)  f_{k_2}^\ast(\eta) 
	\nonumber\\
	&= \frac{H^2}{\sqrt{4k_1^3k_2^3}} \left[k_1^2k_2^2 \eta^2 +\vec{k}_1\cdot\vec{k}_2(1-ik_{12}\eta - k_1k_2\eta^2)\right] e^{ik_{12}\eta}
	\nonumber\\
	&= \frac{H^2}{\sqrt{4k_1^3k_2^3}} \mathcal{O}_B\left(\frac{k_1}{k_3},\frac{k_2}{k_3} \right) e^{ik_{12}\eta},
\end{align}  
where $k_{12}\equiv k_1 + k_2$, and $\mathcal{O}_B$ is the boundary operator given by
\footnote{Our definition of the boundary operator \eqref{def_boundary_operator} differs from the equation (4.53) in \cite{Arkani-Hamed:2015bza} by a negative sign.}
\begin{align}\label{def_boundary_operator}
	\mathcal{O}_B &= \vec{k}_1\cdot\vec{k}_2 - \vec{k}_1\cdot\vec{k}_2\, k_{12} \partial_{k_{12}}
	+ (\vec{k}_1\cdot\vec{k}_2k_1k_2 - k_1^2k_2^2)\partial_{k_{12}}^2 \\\nonumber
	&= \frac{k_3^2}{4} \left[(2-P-Q)(1-P\partial_P) + \frac{1}{2} (P^2-Q^2)(1-P^2) \partial_P^2\right].
\end{align}
Here $2\vec{k}_1\cdot\vec{k}_2 = k_3^2 - k_1^2- k_2^2$ is used and $P$, $Q$ are the two independent variables of the kinematic triangle configuration under momentum conservation. They are given by
\begin{align}
	P \equiv \frac{k_1 + k_2}{k_3}, \qquad Q  \equiv \frac{k_1 - k_2}{k_3}.
\end{align}
We can now pull the boundary operator out of the time-ordering integral to simplify the calculation. As a result, we find
\begin{align}\label{non_time_order_23}
	\int_{t_0}^{t}  d\tilde{t}_1\int_{t_0}^{t} dt_1 &  \langle 0 \vert H_{I2}(\tilde{t}_1)\delta\phi_I^3 H_{I3}(t_1) \vert 0 \rangle^\prime 
	 \nonumber\\
	 &=
	  -c_2c_3\; e^{-2x_\ast} \left(\frac{k_\ast}{H}\right)^{-2\Delta_\epsilon} \frac{\sigma_0^2\dot{\phi}_0}{M_P^4} 
	\frac{H^3}{8k_1^3k_2^3 k_3^2} \mathcal{O}_B\,  \left(\frac{k_3}{H}\right)^{\Delta_\epsilon}\mathcal{S}_{\rm PS}^{23}(P) , \\
	&= -c_2c_3\; e^{-2x_\ast}  \frac{\sigma_0^2\dot{\phi}_0}{M_P^4} 
	\frac{H^3}{8k_1^3k_2^3 k_3^2}   \left(\frac{k_\ast}{H}\right)^{-\Delta_\epsilon} \mathcal{O}_B 
	  \, S_{\rm PS}^{23}(P),
\end{align}
where $S_{\rm PS} = x_{k_3}^{\Delta_\epsilon} \mathcal{S}_{\rm PS} = (k_3/k_\ast)^{\Delta_\epsilon} \mathcal{S}_{\rm PS}$ collects the spectral shape function that is independent of the coupling constants of the vertices. 
In terms of the factorized mode function \eqref{def_B_tilde}, the shape function for the simplest quantum primordial clock is constructed by only one integral with its complex conjugation as
\begin{align}\label{def_INT_PS}
	I_{\rm PS}(\Delta_\epsilon, M_\ast, P) &= \int_{0}^{\infty} d\tau \, \tau^{-1-\frac{\Delta_\epsilon}{2}} e^{i(P+1)\tau} \tilde{B}(\tau), 
	\\\label{def_INT_bar}
	\bar{I}_{\rm PS}(\Delta_\epsilon, M_\ast, P) &=  \int_{0}^{\infty} d\tau \, \tau^{-1-\frac{\Delta_\epsilon}{2}} e^{-i(P+1)\tau} \tilde{B}^\ast(\tau), 
\end{align}
where $\tau = -k_3\eta$. For each time-ordering result,  $\mathcal{S}_{\rm PS}$ is given by
\begin{align}\label{def_SNT23}
	\mathcal{S}_{\rm PS}^{23} &= \frac{\pi}{4} e^{-\mu\pi}\; I_{\rm PS}(\Delta_\epsilon, M_\ast, 1) \times	\bar{I}_{\rm PS}(\Delta_\epsilon, M_\ast, P), 
	\\\label{def_SNT32}
	\mathcal{S}_{\rm PS}^{32} &= \frac{\pi}{4} e^{-\mu\pi}\; I_{\rm PS}(\Delta_\epsilon, M_\ast, P) \times	\bar{I}_{\rm PS}(\Delta_\epsilon, M_\ast, 1).
\end{align}
Here $I_{\rm PS}(\Delta_\epsilon, M_\ast, 1)$ means the result of the integral \eqref{def_INT_PS} with $P = 1$, which corresponds to a ``collapsed'' triangle of the kinematic configuration. $\mathcal{S}_{\rm PS}^{32}  $ denotes the shape function obtained from the three-point anti-time-ordering and the two-point time-ordering correlation, where one can derive similarly to find that
\begin{align}\label{non_time_order_32}
	\int_{t_0}^{t}  d\tilde{t}_1\int_{t_0}^{t} dt_1 &  \langle 0 \vert H_{I3}(\tilde{t}_1)\delta\phi_I^3 H_{I2}(t_1) \vert 0 \rangle^\prime =
	\\\nonumber &
	= -c_2c_3\; e^{-2x_\ast}  \frac{\sigma_0^2\dot{\phi}_0}{M_P^4} 
	\frac{H^3}{8k_1^3k_2^3 k_3^2}   \left(\frac{k_\ast}{H}\right)^{-\Delta_\epsilon} \mathcal{O}_B 
	\, S_{\rm PS}^{32}(P),
\end{align}
and $S_{\rm PS}^{32} = x_{k_3}^{\Delta_\epsilon} \mathcal{S}_{\rm PS}^{32} $.

Results of the slow-roll scalaron approximation can be worked out in a same manner. The model-independent shape function is defined according to
\begin{align}
	\int_{t_0}^{t}d\tilde{t}_1\int_{t_0}^{t} dt_1 &\langle 0 \vert H_I(\tilde{t}_1)\delta\phi_I^3 H_I(t_1) \vert 0 \rangle^\prime =\\\nonumber
	&=  -c_2c_3\; e^{-2x_\ast}  \frac{\sigma_0^2\dot{\phi}_0}{M_P^4} 
	\frac{H^3}{8k_1^3k_2^3 k_3^2}   \left(\frac{k_\ast}{H}\right)^{-\Delta_\epsilon} \mathcal{O}_B 
	\, S_{\rm SR}^{}(P),
\end{align}
with $S_{\rm SR} = S_{\rm SR}^{23} + S_{\rm SR}^{32}$ and $S_{\rm SR} = x_{k_3}^{-\Delta_\epsilon} \mathcal{S}_{\rm SR}$, where
\begin{align}\label{def_S23_SRS}
	\mathcal{S}_{\rm SR}^{23} &= \frac{\pi}{4} e^{-\mu\pi}\; I_{\rm SR}(\Delta_\epsilon, M_\ast, 1) \times	\bar{I}_{\rm SR}(\Delta_\epsilon, M_\ast, P), 
	\\\label{def_S32_SRS}
	\mathcal{S}_{\rm SR}^{32} &= \frac{\pi}{4} e^{-\mu\pi}\; I_{\rm SR}(\Delta_\epsilon, M_\ast, P) \times	\bar{I}_{\rm SR}(\Delta_\epsilon, M_\ast, 1).
\end{align}
The integration of the slow-roll scalaron approximation takes the form of
\begin{align}\label{def_INT_SRS}
	I_{\rm SR}(\Delta_\epsilon, M_\ast, P) &=
	 \int_{0}^{\infty} d\tau \, \tau^{-1+\frac{\Delta_\epsilon}{2}} \left(\frac{55}{\ln\tau - \ln x_k +55}\right) e^{i(P+1)\tau} \tilde{B}(\tau), 
	\\\label{def_INT_SRS_bar}
	\bar{I}_{\rm SR}(\Delta_\epsilon, M_\ast, P) &= 
	\int_{0}^{\infty} d\tau \, \tau^{-1+\frac{\Delta_\epsilon}{2}}  \left(\frac{55}{\ln\tau - \ln x_k +55}\right)  e^{-i(P+1)\tau} \tilde{B}^\ast(\tau),
\end{align}
where $\tilde{B}$ is the numerical solution of \eqref{eom_tilde_B_SR_phi}.

In the standard case ($\Delta_\epsilon = 0$) with the analytic solution $\tilde{B}(\tau)e^{i\tau} = \sqrt{\tau}H_{i\mu}^{(1)}(\tau)$ of the scalar mode function, the integrals \eqref{def_INT_PS} and \eqref{def_INT_bar} can be resolved analytically. The analytic formula is given in Appendix~\ref{Appen_C}. In Figure~\ref{fig.S_non_time_ordered}, we present the numerical results of the model-independent shape function $S_{\rm NT} = S_{\rm NT}^{23} + S_{\rm NT}^{32}$ from the non-time-ordered term \eqref{3pt_nontime_ordered}. The standard analytic predictions \eqref{def_ana_INT} and \eqref{def_ana_INTbar} are in well agreement with the numerical results of $\Delta_\epsilon = 0$ from the pivot-scale expansion approach \eqref{def_INT_PS} and \eqref{def_INT_bar}: 
\begin{align}
	S_{\rm PS}^0 =   \frac{\pi}{4} e^{-\mu\pi} 
	\left\{ I_{\rm PS}(0, M_\ast, 1) \times	\bar{I}_{\rm PS}(0, M_\ast, P)	+  I_{\rm PS}(0, M_\ast, P) \times	\bar{I}_{\rm PS}(0, M_\ast, 1) \right\},
\end{align}
which we can referred as the simplest quantum primordial standard clock.

The standard clock signal cares only the shape of the triangle configuration rather than its length $k_t = k_1 + k_2 + k_3$. However, when the dilatation symmetry is broken by the massive scalar mode function (assigned as $k_3$), the particle exchange process at different scales results in different shape functions. This can be seen from the upper and lower panels in Figure~\ref{fig.S_non_time_ordered} for the shape functions in $R^2$ inflation scaling over two different regimes.

\subsection{The clock amplitude}
\begin{figure}[]
	\begin{center}
		\includegraphics[width= 14 cm]{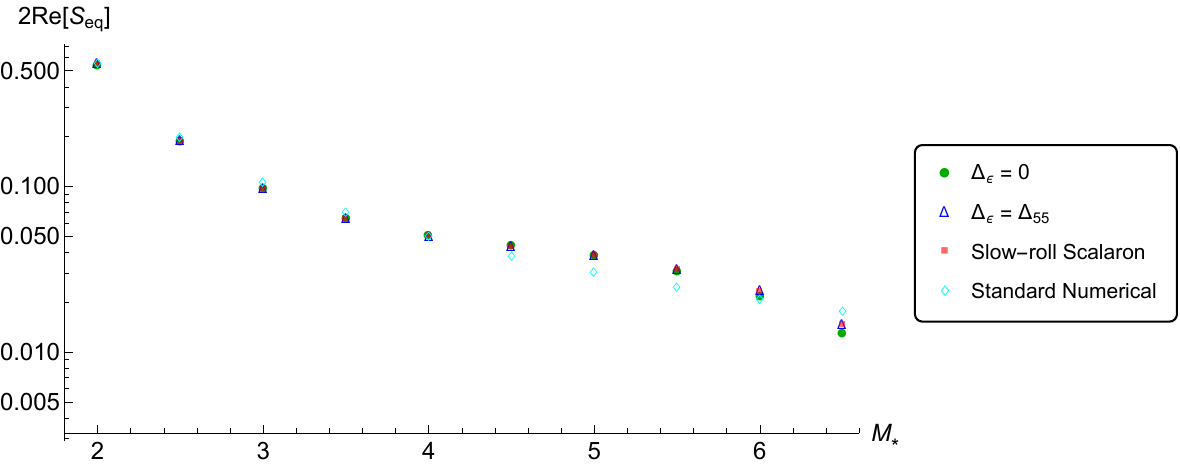}
	\end{center}
	\caption{\label{fig.Seq} 
		The value of the combined shape function given in \eqref{3pt_final_result} at the equilateral limit ($P = 2$) with $x_{k_3} = k_3 /k_\ast =1$.
	}
\end{figure}

We now include the time-ordered contribution \eqref{3pt_time_ordered} to explore the typical size of the entire shape function. The time-order term is involved with two layers of integration. For the pivot-scale expansion appraoch, the time-ordered term is composed as
\begin{align}
	\int_{t_0}^{t}dt_1 & \int_{t_0}^{t_1}dt_2  \langle 0\vert \delta\phi_I^3 H_I(t_1)H_I(t_2)\vert 0\rangle^\prime \\\nonumber
&= \int_{t_0}^{t}dt_1\int_{t_0}^{t_1}dt_2 \left[\langle 0\vert \delta\phi_I^3 H_{I2}(t_1)H_{I3}(t_2)\vert 0\rangle^\prime 
+\langle 0\vert \delta\phi_I^3 H_{I3}(t_1)H_{I2}(t_2)\vert 0\rangle^\prime \right] \\
&= c_2c_3 e^{-2x_\ast} \left(\frac{k_\ast}{H}\right)^{-2\Delta_\epsilon} \frac{\sigma_0^2\dot{\phi}_0}{M_P^4} 
f_{k_1}(0)f_{k_2}(0)f_{k_3}(0) \frac{H^2}{\sqrt{4k_1^3k_2^3}} \mathcal{O}_B 
\label{3pt_time_order_eta_form}\\\nonumber
&\quad\times \left[\int_{-\infty}^{0}d\eta_1\, a(\eta_1)^{3+\Delta_\epsilon} f_{k_3}^{\ast\prime}(\eta_1) u_{k_3}(\eta_1)  \int_{-\infty}^{\eta_1}d\eta_2\, a(\eta_2)^{2+\Delta_\epsilon} e^{ik_{12}\eta_2} u_{k_3}^\ast(\eta_2) \right. \\\nonumber
&\quad \left.+ \int_{-\infty}^{0}d\eta_1 \,a(\eta_1)^{2+\Delta_\epsilon} e^{ik_{12}\eta_1} u_{k_3}(\eta_1)\int_{-\infty}^{\eta_1}d\eta_2 \,a(\eta_2)^{3+\Delta_\epsilon} f_{k_3}^{\ast\prime}(\eta_2) u_{k_3}^{\ast}(\eta_2) \right],
\end{align}
where $k_{12} = k_1 + k_2$ and $\mathcal{O}_B$ is the boundary operator defined in \eqref{def_boundary_operator}.
It is convenient to define the time-ordered shape function $S_{\rm TO,PS} = x_{k_3}^{\Delta_\epsilon}\mathcal{S}_{\rm TO,PS}$ with the definition
\begin{align}\label{def_STO23}
	\mathcal{S}_{\rm TO,PS}^{23}(P) &= \frac{\pi}{4}e^{-\mu\pi} \int_{0}^{\infty} d\tau_1 \tau_1^{-1-\frac{\Delta_\epsilon}{2}}  \tilde{B}(\tau_1) 
	\int_{\tau_1}^{\infty} d\tau_2 \tau_2^{-1-\frac{\Delta_\epsilon}{2}} e^{-i(P+1)\tau_2} \tilde{B}^\ast(\tau_2), \\\label{def_STO32}
	\mathcal{S}_{\rm TO,PS}^{32}(P) &= \frac{\pi}{4}e^{-\mu\pi} \int_{0}^{\infty} d\tau_1 \tau_1^{-1-\frac{\Delta_\epsilon}{2}} e^{-i (P-1) \tau_1} \tilde{B}(\tau_1) 
	\int_{\tau_1}^{\infty} d\tau_2 \tau_2^{-1-\frac{\Delta_\epsilon}{2}} e^{-2i\tau_2} \tilde{B}^\ast(\tau_2), 
\end{align} 
and $\tilde{B}$ is the numerical solution of \eqref{eom_tilde_B}.
As a result, we can rewrite \eqref{3pt_time_order_eta_form} as
\begin{align}
\int_{t_0}^{t}dt_1  \int_{t_0}^{t_1}dt_2  & \langle 0\vert \delta\phi_I^3 H_I(t_1)H_I(t_2)\vert 0\rangle^\prime	\\\nonumber
&=-c_2c_3\; e^{-2x_\ast}  \frac{\sigma_0^2\dot{\phi}_0}{M_P^4} 
\frac{H^3}{8k_1^3k_2^3 k_3^2}   \left(\frac{k_\ast}{H}\right)^{-\Delta_\epsilon} \mathcal{O}_B \;
 2\Re\left[S_{\rm TO}^{23} + S_{\rm TO}^{32}\right].
\end{align}
Meanwhile, we can also rewrite the non-time-ordered term \eqref{3pt_nontime_ordered} in a similar way as
\begin{align}
	\int_{t_0}^{t}dt_1\int_{t_0}^{t} dt_2 &\langle 0 \vert H_I(t_1)\delta\phi_I^3 H_I(t_2) \vert 0 \rangle^\prime \\\nonumber
	&=-c_2c_3\; e^{-2x_\ast}  \frac{\sigma_0^2\dot{\phi}_0}{M_P^4} 
	\frac{H^3}{8k_1^3k_2^3 k_3^2}   \left(\frac{k_\ast}{H}\right)^{-\Delta_\epsilon} \mathcal{O}_B \;
	 \left[S_{\rm NT}^{23} + S_{\rm NT}^{32}\right],
\end{align}
where we can reexpress the shape functions $S_{\rm NT,PS} = x_{k_3}^{\Delta_\epsilon}\mathcal{S}_{\rm NT,PS}$ as
\begin{align}
	\mathcal{S}_{\rm NT,PS}^{23}(P) &=\frac{\pi}{4}e^{-\mu\pi} \int_{0}^{\infty} d\tau_1 \tau_1^{-1-\frac{\Delta_\epsilon}{2}} e^{2i\tau_1} \tilde{B}(\tau_1) 
	\int_{0}^{\infty} d\tau_2 \tau_2^{-1-\frac{\Delta_\epsilon}{2}} e^{-i(P+1)\tau_2} \tilde{B}^\ast(\tau_2), \\
	\mathcal{S}_{\rm NT,PS}^{32}(P) &= \frac{\pi}{4}e^{-\mu\pi} \int_{0}^{\infty} d\tau_1 \tau_1^{-1-\frac{\Delta_\epsilon}{2}} e^{i (P+1) \tau_1} \tilde{B}(\tau_1) 
	\int_{0}^{\infty} d\tau_2 \tau_2^{-1-\frac{\Delta_\epsilon}{2}} e^{-2i\tau_2} \tilde{B}^\ast(\tau_2). 
\end{align}
The structure difference between $\mathcal{S}_{\rm NT}$ and $\mathcal{S}_{\rm TO}$ given above is very similar to the case of the two-point correlation function. Therefore we perform the same trick as used in \eqref{def_C} by decomposing the non-time-ordered integral as $\int_{0}^{\infty}d\tau_1\int_{0}^{\infty}d\tau_2 = \int_{0}^{\infty}d\tau_1 (\int_{\tau_1}^{\infty}+\int_{0}^{\tau_1})d\tau_2$. After exchanging the order of integration to the second term, we find
\begin{align}
	\mathcal{S}_{\rm NT,PS}^{23} &+\mathcal{S}_{\rm NT,PS}^{32} = 2\Re\left[\mathcal{S}_{\rm NT,PS}^{23}\right] \\\nonumber
	&= \frac{\pi}{4}e^{-\mu\pi} 2\Re\left[ \int_{0}^{\infty} d\tau_1 \tau_1^{-1-\frac{\Delta_\epsilon}{2}} e^{2i\tau_1} \tilde{B}(\tau_1) 
	\int_{\tau_1}^{\infty} d\tau_2 \tau_2^{-1-\frac{\Delta_\epsilon}{2}} e^{-i(P+1)\tau_2} \tilde{B}^\ast(\tau_2)\right]\\\nonumber
	&\quad+ \frac{\pi}{4}e^{-\mu\pi} 2\Re\left[ \int_{0}^{\infty} d\tau_1 \tau_1^{-1-\frac{\Delta_\epsilon}{2}} e^{i (P+1) \tau_1} \tilde{B}(\tau_1) 
	\int_{\tau_1}^{\infty} d\tau_2\tau_2^{-1-\frac{\Delta_\epsilon}{2}} e^{-2i\tau_2} \tilde{B}^\ast(\tau_2) \right].
\end{align}
This result allows us to write down a unified shape function as
\begin{align}
	\mathcal{S}_{\rm NT}^{23} +\mathcal{S}_{\rm NT}^{32} -2\Re\left[\mathcal{S}_{\rm TO}^{23} +\mathcal{S}_{\rm TO}^{32}\right] 
	= 2\Re\left[\mathcal{S}_{\rm mix}^{23} +\mathcal{S}_{\rm mix}^{32}\right],
\end{align}
where $\mathcal{S}_{\rm mix}$ is the combined representation of the time-ordered and non-time-ordered contributions given by
\begin{align}
	\mathcal{S}_{\rm mix, PS}^{23}(P) &= \frac{\pi}{4}e^{-\mu\pi} 
	\int_{0}^{\infty} d\tau_1 \tau_1^{-1-\frac{\Delta_\epsilon}{2}} \left(e^{2i\tau_1} - 1\right) \tilde{B}(\tau_1) 
	\int_{\tau_1}^{\infty} d\tau_2 \tau_2^{-1-\frac{\Delta_\epsilon}{2}} e^{-i(P+1)\tau_2} \tilde{B}^\ast(\tau_2), \\
	\mathcal{S}_{\rm mix,PS}^{32}(P) &= \frac{\pi}{4}e^{-\mu\pi} 
	\int_{0}^{\infty} d\tau_1 \tau_1^{-1-\frac{\Delta_\epsilon}{2}} \left(e^{i (P+1) \tau_1} - e^{-i (P-1) \tau_1}\right) \tilde{B}(\tau_1) 
	\nonumber\\
	&\;\qquad\qquad\times   \int_{\tau_1}^{\infty} d\tau_2 \tau_2^{-1-\frac{\Delta_\epsilon}{2}} e^{-2i\tau_2} \tilde{B}^\ast(\tau_2). 
\end{align}
It is not difficult to see that the combination of $e^{i(P+1)\tau} -e^{-i(P-1)\tau}$, with $P \geq 1$ being a real number, vanishes in the limit of $\tau\rightarrow 0$, which regularized the behavior of the shape function in the late-time limit even with $\Delta_\epsilon > 0$. 

Again, we can repeat the procedure for the slow-roll scalaron approximation to obtain the combined shape functions $S_{\rm mix,SR} = x_k^{-\Delta_\epsilon} \mathcal{S}_{\rm mix,SR}$:
\begin{align}
	\mathcal{S}_{\rm mix,SR}^{23}(P) &= \frac{\pi}{4}e^{-\mu\pi} 
	\int_{0}^{\infty} d\tau_1 \;\tau_1^{-1+\frac{\Delta_\epsilon}{2}} \left(\frac{55}{\ln\tau_1 -\ln x_k +55}\right) \left(e^{2i\tau_1} - 1\right) \tilde{B}(\tau_1)  
	\nonumber\\
	&\qquad\times \int_{\tau_1}^{\infty} d\tau_2\; \tau_2^{-1+\frac{\Delta_\epsilon}{2}}\left(\frac{55}{\ln\tau_2 -\ln x_k +55}\right)  e^{-i(P+1)\tau_2} \tilde{B}^\ast(\tau_2), \\
	\mathcal{S}_{\rm mix,SR}^{32}(P) &= \frac{\pi}{4}e^{-\mu\pi} 
	\int_{0}^{\infty} d\tau_1\; \tau_1^{-1+\frac{\Delta_\epsilon}{2}} \left(\frac{55}{\ln\tau_1 -\ln x_k +55}\right)  \left(e^{i (P+1) \tau_1} - e^{-i (P-1) \tau_1}\right) \tilde{B}(\tau_1) 
	\nonumber\\
	&\qquad\times\int_{\tau_1}^{\infty} d\tau_2\; \tau_2^{-1+\frac{\Delta_\epsilon}{2}} \left(\frac{55}{\ln\tau_2 -\ln x_k +55}\right)  e^{-2i\tau_2} \tilde{B}^\ast(\tau_2). 
\end{align}
In summary, the three-point correlation function due to the massive scalar exchange process computed in this section reads
\begin{align}\label{3pt_final_result}
	\left\langle \delta\phi^3\right\rangle^\prime =
	-c_2c_3\; e^{-2x_\ast}  \frac{\sigma_0^2\dot{\phi}_0}{M_P^4} 
	\frac{H^3}{8k_1^3k_2^3 k_3^2}   \left(\frac{k_\ast}{H}\right)^{-\Delta_\epsilon} \mathcal{O}_B \; 2\Re\left[S_{\rm mix}\right],
\end{align}
where $S_{\rm mix} = S_{\rm mix}^{23} + S_{\rm mix}^{32}$ and $S_{\rm mix,PS} = x_{k3}^{\Delta_\epsilon}\mathcal{S}_{\rm mix,PS}$ for solutions from the pivot-scale expansion approach \eqref{eom_tilde_B} while $S_{\rm mix,SR} = x_{k3}^{-\Delta_\epsilon}\mathcal{S}_{\rm mix,SR}$ for solutions from the slow-roll scalaron approximation \eqref{eom_tilde_B_SR_phi}.

In Figure~\ref{fig.Seq}, we perform a scan of the amplitude for the combined shape function $\mathcal{S}_{\rm mix}$ at the equilateral limit $P = 2$ with $k_1 = k_2 = k_3 = k_\ast$. The amplitude of the shape function runs from $\mathcal{O}(0.5)$ with $M_\ast = 2$ to $\mathcal{O}(10^{-2})$ with $M_\ast = 6$. For $M_\ast < 6$, the numerical errors does not lead to significant deviations to the values obtained from the numerical computations based on the standard analytic mode functions with exact dilatation invariance.
 
At the end of this section, let us check the typical size of the non-Gaussianity relevant to the simplest quantum primordial clock. We adopt the definition of the shape function from the curvature perturbation as \cite{Chen:2010xka}:
\begin{align}
	\left\langle \zeta^3\right\rangle = \left(-\frac{H}{\dot{\phi}_0}\right)^3 \langle\delta\phi^3\rangle 
	=(2\pi)^7 \delta^{(3)}(\vec{k}_1+\vec{k}_2+\vec{k}_3) B(k_1,k_2,k_3) \frac{A_\zeta^2}{(k_1k_2k_3)^2},
\end{align} 
where $(2\pi)^2A_\zeta = H^4/\dot{\phi}_0^2 \approx 2.2\times 10^{-9}$ is the amplitude of the power spectrum of the curvature perturbation $\zeta$. This gives the definition of $f_{\rm NL}$ in the equilateral limit as
\begin{align}
	\left\langle \zeta^3 \right\rangle \rightarrow (2\pi)^7 \delta^{(3)}\left(\vec{k}_1+\vec{k}_2+\vec{k}_3\right) 
	\frac{A_\zeta^2}{k_\ast^6} \left(\frac{9}{10} f_{\rm NL}^{\rm eq}\right),
\end{align}
where we have used $k_1 = k_2 = k_3 = k_\ast$.
If we take $c_2 = c_3 = 1$ and pick up $\Delta_\epsilon = \Delta_{55}$, $\epsilon_U = 2.6 \times 10^{-4}$, $\phi_\ast/M_P = 5.26$ based on $\Delta N = 55$ in $R^2$ inflation, we find that at the zeroth order of the operator $\mathcal{O}_B \sim \vec{k}_1\cdot\vec{k}_2 = -k_\ast^2/2$, and the result \eqref{3pt_final_result} for both numerical solutions lead to the similar estimation
\begin{align}
	 f_{\rm NL}^{\rm eq} \lesssim 10^{-8}\times \frac{\sigma_0^2}{M_P^2} \times \left(\frac{k_\ast}{H}\right)^{-\Delta_{55}} \times 2\Re\left[S_{\rm mix}^{\rm eq}\right], 
\end{align}
where the maximal value $(k_\ast/H)^{-\Delta_{55}}\approx 10$ is given by the highest inflationary Hubble scale $H = 3\times 10^{13}$ GeV according to \eqref{enhancement_factor}. The amplitude of $S_{\rm mix}^{\rm eq}$ is shown in Figure~\ref{fig.Seq}.

\section{Conclusions and discussions}\label{Sec. conclusion}
In this work we have investigated numerical solutions of the quantum mode functions of a massive scalar field during $R^2$ inflation, focusing on the time-varying mass effect led by the conformal coupling with the rolling scalaron. 
We have examined the tree-level corrections to the primordial power spectrum and the simplest realization of the so-called quantum primordial clock signals induced by the non-local propagation of the massive scalar perturbations under the slow-roll background created by the $R^2$ model. These tree-level corrections can occur with a small but non-vanishing homogeneous (or the zero-mode) motion of the massive scalar near a local potential minimum. As shown in Appendix~\ref{Appen_mixed_Higgs_R2}, the  isocurvature scalar motion can be supported by a non-minimal coupling with gravity while keeping a negligible backreaction to the scalaron motion along the standard $R^2$ inflationary trajectory. 
The most important findings of this work can be outlined as:
\begin{itemize}
	\item The slow-roll inflaton $\phi$ (or namely the scalaron) in the Einstein frame of $R^2$ inflation couples to all matter fields through the unique factor $e^{-\alpha\phi/M_P}$ with $\alpha = \sqrt{2/3}$ under the conformal transformation, leading to the breaking of dilatation symmetry in the mode functions of matter perturbations. For a massive scalar field $\sigma$, the slow-roll scalaron results in a time-varying mass $m^2(t) \sim e^{-\alpha\phi(t)/M_P}m_\sigma^2$ and a modified decay rate $\delta\sigma\sim a^{-(3+\Delta_\epsilon)/2}$ on superhorizon scales, where $\Delta_\epsilon \approx 0.018$ \eqref{def_Delta55}.
	
	\item For a pivot scale $k_\ast = 0.002$ Mpc$^{-1}$, the modified decay rate for a massive scalar perturbation $\delta\sigma$ results in a universal enhancement $(k_\ast/H)^{-\Delta_\epsilon} \lesssim 10$ to all primordial correlators involved with the exchange process for $\delta\sigma$. The maximal enhancement is given by the highest inflationary Hubble scale $H = 3\times 10^{13}$ GeV \eqref{enhancement_factor} from the recent bound \cite{BICEP:2021xfz}. For the corrections to the primordial power spectrum, we find
	\begin{align}
		\frac{\Delta P}{P_0} \sim \beta_{\rm model} \times C(\Delta_\epsilon, M_\ast, k/k_\ast),
	\end{align}  
	where $\beta_{\rm model} \sim 10^{-8}\times \sigma_0^2/M_P^2$ depends on the models of the massive scalar $\sigma$, $\sigma_0$ is the scalar VEV, and $P_0$ is the scalaron power spectrum without corrections. $\Delta P$ is the contribution from the vertex \eqref{def_L2} only. The numerical factor $C$ is given by \eqref{def_C_B_tilde} or \eqref{def_C_B_tilde_SRS}.
	
	
	\item The explicit scale and time dependence in the mode functions of the scalar perturbations brings in scale dependence to the shape functions of primordial bispectra. Figure~\ref{fig.S_non_time_ordered} shows the discrepancy of the spectral shape functions in the same (equilateral-to-squeezed) configuration but scaling over two different ranges of $k$-scales. The typical size of non-Gaussianity corresponding to the exchange process led by \eqref{def_L2} and \eqref{def_L3} is
	\begin{align}
		f_{\rm NL}^{\rm eq} \sim \beta_{\rm model} \times \left(\frac{k_\ast}{H}\right)^{-\Delta_\epsilon} \times 2\Re\left[S_{\rm mix}^{\rm eq}\right],
	\end{align} 
	with the same $\beta_{\rm model}$ from the power spectrum and $2\Re[S_{\rm mix}^{\rm eq}] \lesssim 0.5$ is shown in Figure~\ref{fig.Seq}.
\end{itemize} 

Given that we only consider vertices arisen from the conformal factor via integration by parts, each vertex of \eqref{def_L2} or \eqref{def_L3} is at least suppressed by $M_P^{-2}$. This already makes all the relevant predictions very difficult to be tested by future experiments. Futhermore, the total tree-level contribution in the standard $R^2$ inflationary background is suppressed by the smallness of the model-dependent zero-mode motion of the isocurvature scalar, and in general we expect leading corrections come from loops if the scalar is completely frozen at a local minimum.   Nevertheless, the investigation is still worth an effort, since $R^2$ inflation is the best-fit scenario. Moreover, we have not yet explored the full possibilities to obtain some much larger $\beta_{\rm model}$ from different model buildings. Following the findings of this work, there are several topics remaining to be studied. For example:
\begin{itemize}
	\item \textit{Higgs inflation.} In the large-field regime ($\phi/M_P \gg 1$), the conformal factor in the Higgs inflation with a non-minimal coupling to gravity \cite{Bezrukov:2007ep} coincides with that in the $R^2$ model (see Appendix~\ref{Appen_Higgs_inflation}). In Higgs inflation, one can write down direct couplings of Higgs with scalar fields, for example, in the form of $ h^2\sigma^2 \sim \frac{M_P^2}{\xi} e^{\alpha\phi/M_P} \sigma^2$. Direct couplings of the inflaton with scalar fields can lead to unsuppressed interaction vertices. Of course, the time dependence of the scalar mass could be changed drastically by the introduction of such a direct coupling.
	
	\item \textit{Particles with spins.} The characteristic signals for spin $1/2$ fermions and gauge bosons are also important targets for the cosmological collider program \cite{Wang:2019gbi,Wang:2020ioa}. The Lagrangian which describes the dynamics of these matter fields with non-zero spins must include covariant derivatives with respect  to the spacetime metric. Moving to the Einstein frame for the inflaton via the conformal transformation, those terms involved with covariant derivatives will partially cancel out the conformal factor $\Omega$, leading to a breaking of the dilatation invariance in the equation of motion as the case for a real scalar field investigated in this work.
	
	\item \textit{The Standard Model mass spectrum.} The conformal coupling acts on all matter fields, including the Standard Model particles. The clarification of the mass spectrum of the Standard Model is the important first step towards the discovery of new particles by using the cosmological collider \cite{Chen:2016hrz,Chen:2016uwp}. If $R^2$ inflation is the correct model, the Standard Model predictions shall be modified by the effects considered in this work. 
\end{itemize} 

We leave the further researches of these interesting topics as future efforts.

\acknowledgments
The author mourns the loss of professor Alexei Starobinsky during the construction of this project.
We would like to thank Daniel Baumann, Xingang Chen, Soubhik Kumar, Enrico Pajer, Guilherme Pimentel, Santiago Salcedo, Zhong-Zhi Xianyu for many helpful discussions, and we are grateful to the technical support from Wei-Xiang Feng, Yen-Hsun Lin, Hung Tan, and Yi Wang. 
Y.-P. Wu acknowledges the Bethe Forum ``Inflation'' at the Bethe Center for Theoretical Physics for many inspiring ideas of this work.
The project has in part received funding from the European Union’s Horizon 2020 research and innovation programme under grant agreement No 101002846 (ERC CoG “CosmoChart”).

\appendix
\section{Conformal couplings in models of inflation}\label{Appen_A}
In this section we outline the matter couplings with the canonically normalized inflaton $\phi$ in the Einstein frame of inflationary models originated from the conformal transformation of the metric as
\begin{align}\label{metric_conformal_trans}
	g_{\mu\nu}^E = \Omega^2 \; g_{\mu\nu}^J,
\end{align} 
where $\Omega$ denotes the conformal factor and $g_{\mu\nu}^E$ ($g_{\mu\nu}^J$) corresponds to the metric tensor in the Einstein (Jordan) frame. The Ricci scalars in the two frames are related by
\begin{align}\label{Ricci_relation}
	R_J = \Omega^2 \left[R_E + 6\square_E \ln\Omega -6 g_E^{\mu\nu}\partial_\mu \ln\Omega\; \partial_\nu \ln\Omega \right],
\end{align}
where the combination $\sqrt{-g_E}\square_E  \ln\Omega $ becomes a total covariant derivative in the Einstein frame in all cases studied in this section.
Note that $\sqrt{-g_E} = \Omega^4 \sqrt{-g_J}$.

\subsection{$R^2$ inflation and $f(R)$ gravity}\label{Append_R2}
For the discussion of the $R^2$ model, it is convenient to start with the more general $f(R)$ gravity so that we can reformulate the theory into a scalar-tensor action via an auxiliary scalar field $\chi$ \cite{Sotiriou:2008rp,DeFelice:2010aj}. The Jordan frame action transforms as
\begin{align}\label{fR_Jordan}
S= 	\frac{M_P^2}{2} \int d^4x \sqrt{-g_J} \; f(R_J) \equiv \frac{M_P^2}{2} \int d^4x \sqrt{-g_J} \left[f(\chi) +(R_J -\chi)f_\chi(\chi) \right],
\end{align}
where $f_\chi \equiv \partial_\chi f$ and the variation with respect to $\chi$ for the second expression gives nothing but 
\begin{align}
	f_{\chi\chi} (R_J - \chi) =0.
\end{align}
This implies that $\chi = R_J$ as long as $f_{\chi\chi} \neq 0$.

We can now assign the scalar degree of freedom in terms of the conformal factor as $f_\chi(\chi) \equiv \Omega^2$. This allows us to construct expressions in terms of the conformal factor, such as $\chi = \chi(\Omega)$ and $f(\chi) = f(\chi(\Omega))$. According to \eqref{Ricci_relation} from the conformal transformation \eqref{metric_conformal_trans}, the Einstein frame expression of the action \eqref{fR_Jordan} is given by
\begin{align}
	S = 	\frac{M_P^2}{2} \int d^4x \sqrt{-g_E} \left[R_E -6\left(\partial\ln\Omega\right)_E^2 + \Omega^{-4} W(\Omega)\right],
\end{align}
where $W(\Omega) \equiv f(\chi(\Omega)) -\chi(\Omega) \Omega^2$ can be considered as an effective potential.
It becomes evident that the canonical scalar $\phi$ appears in this frame with the definition
\begin{align}
	\ln\Omega \equiv \frac{1}{\sqrt{6}} \frac{\phi}{M_P}.
\end{align}
As a result for the matter sector, the conformal transformation invokes
\begin{align}
	S_{\rm matter} = \int d^4x \sqrt{-g_E}\; \Omega^{-4} \mathcal{L}_{\rm matter} = \int d^4x \sqrt{-g_E}\; e^{-2\sqrt{\frac{2}{3}} \frac{\phi}{M_P}} \mathcal{L}_{\rm matter}.
\end{align}
Note that it is a generic expression for the $f(R)$ gravity theories.

\begin{figure}[]
	\begin{center}
		\includegraphics[width= 7 cm]{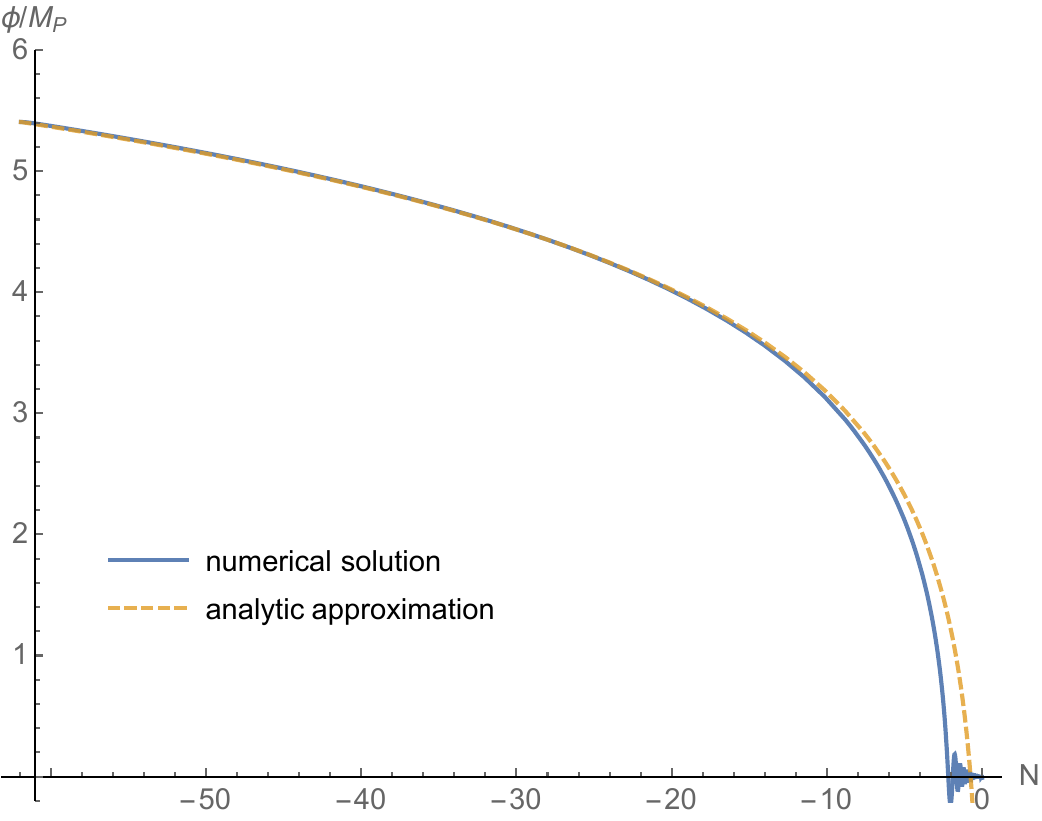}
	\end{center}
	\caption{\label{fig.slowroll_scalaron} 
		The evolution of the scalaron $\phi$ as function of the $e$-folding number $N$ in $R^2$ inflation based on the potential \eqref{U_phi}. The solid line shows the numerical solution from field equations without imposing any slow-roll conditions. The dashed line is the analytic approximation given by \eqref{def_e_fold_phi}.		
	}
\end{figure}

For the $R^2$ model of inflation \eqref{Action_J_frame}, we have the specific form
\begin{align}
	f(\chi) = \chi +\frac{\chi^2}{6M^2}, \qquad \chi = 3M^2 (\Omega^2 - 1).
\end{align} 
This leads to the slow-roll potential of $\phi$ in the Einstein frame as \eqref{U_phi} according to
\begin{align}
	U(\phi) = - \frac{M_P^2}{2} \Omega^{-4} W(\Omega) = \frac{3}{4} M_P^2 M^2 \left( 1- \Omega^{-2}\right)^2,
\end{align}
with $\Omega(\phi) = e^{\phi/(\sqrt{6}M_P)}$. In terms of the dimensionless parameter $x \equiv \sqrt{\frac{2}{3}}\frac{\phi}{M_P}$, one can obtain the slow-roll parameters from the potential 
\begin{align}\label{def_SR_parameters}
	\epsilon_U &\equiv \frac{M_P^2}{2}\left(\frac{U_\phi}{U}\right)^2 &&= \frac{4}{3} \frac{1}{(e^x - 1)^2}, \\
	\eta_U  &\equiv M_P^2\frac{U_{\phi\phi}}{U} &&= \frac{4}{3} \frac{e^{-x}(2e^{-x} - 1)}{(1-e^{-x})^2}.
\end{align}
For a given value of $\phi$, the corresponding e-folding number $N = \ln a$ to the end of inflation can be estimated from \cite{Maggiore:2018sht}:
\begin{align}
	\Delta N \equiv \vert N - N_{\rm end} \vert \approx \frac{1}{M_P^2} \int_{\phi_{\rm end}}^{\phi} d\phi_1 \frac{U(\phi_1)}{U_\phi(\phi_1)}
	\approx \frac{3}{4} e^{\sqrt{\frac{2}{3}} \frac{\phi}{M_P}}.
\end{align}
This estimation can be translated into the useful relation 
\begin{align}\label{slow_roll_phi_analytic}
	\frac{\phi}{M_P} = \sqrt{\frac{3}{2}} \ln\left(\frac{4}{3}\Delta N\right).
\end{align}
In Figure~\ref{fig.slowroll_scalaron}, we compare this analytic approximation with the numerical solution of the scalaron dynamics $\phi(\Delta N)$ in the potential \eqref{U_phi} without imposing any slow-roll conditions.  Assuming that the slow-roll phase ends around $\epsilon_U = 1$, which indicates $\phi_{\rm end}/M_P \sim 1$, we confirm that \eqref{slow_roll_phi_analytic} is a good approximation for the epoch of inflation ($-60 < N < -40$) considered in this work.

\subsection{Higgs inflation with a non-minimal coupling to gravity}\label{Appen_Higgs_inflation}
Let us also take a look at the inflationary model driven by  the Standard Model Higgs boson with a non-minimal coupling $\xi \geq 0$ to the Ricci scalar \cite{Bezrukov:2007ep}. In the Jordan frame, the action of this model is given by
\begin{align}\label{Action_Higgs_inflation}
	S = \int d^4x \sqrt{-g_J} \left[\frac{M_P^2}{2} \left(1+\frac{\xi h^2}{M_P^2}\right)R_J  -\frac{1}{2}\left(\partial h\right)_J^2 - \frac{\lambda}{4} h^4 \right],
\end{align}
where $h$ is the real scalar mode of the Higgs field in the unitary gauge and we drop the electroweak vacuum expectation value since it is irrelevant to our discussion.

Similarly, we shall assign the conformal factor with respect to the scalar function
\begin{align}
	\Omega^2 \equiv 1 + \xi \frac{h^2}{M_P^2},
\end{align}
to obtain the Einstein frame representation of the theory \eqref{Action_Higgs_inflation}, which is given by
\begin{align}
	S = \int d^4x \sqrt{-g_E} \left[\frac{M_P^2}{2} R_E - 6 \left(\partial \ln \Omega\right)_E^2 - \frac{1}{2\Omega^2} \left(\partial h\right)_E^2 -\frac{\lambda}{4\Omega^4} h^4 \right].
\end{align}
Note that the total derivative term in \eqref{Ricci_relation} has been dropped. The canonical inflaton in the Einstein frame can be derived from \cite{Bezrukov:2007ep}:
\begin{align}
	\frac{d \phi}{d h} = \left(\frac{1}{\Omega^2} + \frac{6\xi^2}{\Omega^4}\frac{h^2}{M_P^2}\right)^{1/2}.
\end{align} 
In the small-field limit $\sqrt{\xi}h/M_P \ll 1$, we have $\Omega^2 \approx 1$ so that the Einstein frame action coincides with that of the original Jordan frame. 
However, the slow-roll plateau for the potential of $\phi$ appears in the large-field regime $\sqrt{\xi} h/ M_P \gg 1$, where 
\begin{align}
	\frac{\phi}{M_P} \approx \sqrt{6}  \ln \frac{h}{h_0},
\end{align}
with $h_0 = M_P/\sqrt{\xi}$ showing the cutoff for such a large-field approximation. 
The inflaton potential is led by
\begin{align}
	\frac{\lambda}{4 \Omega^4(h(\phi))} h^4(\phi) \approx \frac{\lambda M_P^4}{4\xi^2} \left(1+ e^{-\sqrt{\frac{2}{3}}\frac{\phi}{M_P} }\right)^{-2}.
\end{align}
Meanwhile, to the leading order in the large-field expansion the matter sector receives the coupling with $\phi$ as
\begin{align}
	S_{\rm matter} = \int d^4x \sqrt{-g_E}\; \Omega^{-4} \mathcal{L}_{\rm matter} 
	\approx \int d^4x \sqrt{-g_E}\; e^{-2\sqrt{\frac{2}{3}} \frac{\phi}{M_P}} \mathcal{L}_{\rm matter},
\end{align}
which takes the same form as in \eqref{Action_E_frame} for the $R^2$ inflation model.

\subsection{The mixed Higgs-$R^2$ inflation}\label{Appen_mixed_Higgs_R2}
In this section we seek for an explicit model that can admit the tree-level processes considered in Figure~\ref{fig.Feynman_clock} from the conformal coupling with the scalaron in the Einstein frame. The presence of a zero-mode motion $\dot{\sigma}_0 \neq 0$ is the key to have the two-point transfer vertex from the kinetic term. A possible realization of this condition can be found in the so-called mixed Higgs-$R^2$ inflation  \cite{He:2018gyf,Ema:2023dxm}, where we start with a general potential for the isocurvature scalar as
\begin{align}
	\label{Action_J_frame_mixed}
	S &= \int d^4 x \sqrt{-g_J}\, \frac{M_P^2}{2} \left[R_J +\frac{R_J^2}{6M^2} +\xi\frac{\sigma^2}{M_P^2}R_J\right] 
	+\int d^4x \sqrt{-g_{J}}\, \left[-\frac{1}{2} (\partial\sigma)_J^2 -V(\sigma)\right], 
	\\\label{Action_E_frame_mixed}
	&= \int d^4x \sqrt{-g_E}\, 
	\left[\frac{M_P^2}{2} R_E - \frac{1}{2}(\partial\phi)_E^2  - \frac{1}{2}e^{-\sqrt{\frac{2}{3}}\frac{\phi}{M_P}}(\partial\sigma)_E^2 -U(\phi, \sigma)\right].
\end{align}
Here $\sigma$ is a real scalar that has a non-minimal coupling $\xi$ with gravity in the Jordan frame \eqref{Action_J_frame_mixed}. In terms of $x = \sqrt{\frac{2}{3}}\frac{\phi}{M_P}$, the potential in the Einstein frame is given by
\begin{align}\label{U_phi_mixed}
	U(x, \sigma) = \frac{3}{4} M_P^2M^2 e^{-2x} \left(e^x - 1- \xi \frac{\sigma^2}{M_P^2}\right)^2 + e^{-2x} V(\sigma),
\end{align}
and the definition of the scalaron reads
\begin{align}
	\sqrt{\frac{2}{3}}\frac{\phi}{M_P} = \ln \left(1 + \xi \frac{\sigma^2}{M_P^2} +\frac{\chi}{3M^2}\right),
\end{align}
where $\chi = R_J$ is the auxiliary scalar field used in \eqref{fR_Jordan}.
Note that $\sigma$ becomes a part of the scalaron if $\xi \neq 0$, and thus we shall require $\xi \sigma^2 /M_P^2 \ll 1$ to avoid a notable deviation from the solution \eqref{slow_roll_phi_analytic} in the single-field inflation limit.

Let us specify the field potential $V(\sigma)$ by a double-well shape as
\begin{align}
	V(\sigma) = \frac{\lambda}{4} \left(\sigma^2 - v_\sigma^2\right)^2,
\end{align}
with a constant barrier $v_\sigma$. The prototype model of the mixed Higgs-$R^2$ inflation \cite{He:2018gyf} can be recovered with $v_\sigma = 0$. 
In the case with $v_\sigma = 0$, there is a special relation between $\xi$ and $M$ which ensures that the power spectrum of the curvature perturbation is effectively the same as the prediction of the standard $R^2$ inflation with $\xi = 0$ \cite{He:2018gyf}. In particular, in this relation $M$ approaches to a constant value when $\xi < 10^3$, which is the parameter space of our main focus. 

For $v_\sigma >0$, the minima of the total potential $U(\phi, \sigma)$ along the $\sigma$ direction at $\partial_\sigma U = 0$ are given by
\begin{align}
	\sigma_{\rm min}^2 = \frac{\lambda v_\sigma^2 + 3\xi M^2 (e^x - 1)}{3\xi^2 M^2/M_P^2 + \lambda}.
\end{align}
If we take $\lambda M_P^2 \gg 3\xi^2 M^2$ and $\lambda v_\sigma^2 \gg 3\xi M^2 (e^x - 1) \approx 3\xi M^2 e^x$, then the isocurvature scalar can have a nearly constant VEV at $\sigma_{\rm min}^2 \approx v_\sigma^2$. This leads to the time-varying mass $U_{\sigma\sigma} \approx 2e^{-2x}\lambda v_\sigma^2$ as we want to solve in \eqref{eom_sigma_general}, and the mass parameter can be identified as $2\lambda v_\sigma^2 = m_\sigma^2$.  Note that $x_\ast \gg 1$ near the pivot scale.
On the other hand, if $\lambda v_\sigma^2 \ll 3\xi M^2 e^x$, the scalar mass in the equation of motion \eqref{eom_sigma_general} $e^x U_{\sigma\sigma} \approx 6\xi M^2$ is a constant. This is the case considered in \cite{Ema:2023dxm}.

For a constant mass $m_\sigma^2 = 2\lambda v_\sigma^2 \gg 3\xi M^2 e^x$, we are mostly interested in the value around $m_\ast^2 = e^{-x}m_\sigma^2 \sim H^2$ for having a mild suppression of the quantum clock signals. This further implies the upper bound of the non-minimal coupling as
\begin{align}
	\xi \ll \frac{H^2}{3M^2} \approx \frac{1}{12},
\end{align}
where the slow-roll assumption $3M_P^2H^2 \approx U(\phi, \sigma) \approx \frac{3}{4}M_P^2 M^2$ is used in the second approximation. Combining the condition $\xi \sigma^2 \ll M_P^2$ for the scalaron in the single-field inflation limit, we also obtain a constraint for the potential parameter as 
\begin{align}
	\frac{M_P^2}{\xi} \gg v_\sigma^2 \gg 3\frac{\xi M^2}{\lambda} e^{x_\ast}.
\end{align} 

Finally, to realize a non-zero $\dot{\sigma}_0$ across the pivot scale, the initial condition of the field value $\sigma$ shall be very close to, but not exactly at, the minima $\sigma_{\rm min}^2 = v_\sigma^2$. This can lead to a slow-roll motion $(3+\Delta_\epsilon)H \dot{\sigma} \approx 3H\dot{\sigma}\approx - e^x \partial_\sigma U$. Since $\partial_\sigma U \vert_{\sigma = v_\sigma} \approx -3\xi M^2 v_\sigma e^{-x}$, we arrive at the relation
\begin{align}\label{suppression_factor_xi}
	\frac{\dot{\sigma}_0}{H v_\sigma} \approx \xi \frac{M^2}{H^2} \approx 4\xi \ll \frac{1}{3}.
\end{align}
In other words, a non-zero $\dot{\sigma}_0$ can be supported by the presence of a small non-minimal coupling $0< \xi \ll 1/12$.

\section{Numerical methodology}\label{Appen_B}
Let us address the initial conditions in the early-time limit ($\tau \gg 1$) for the factorized mode functions \eqref{def_B_tilde_initial_condition} in this section. In the case with dilatation invariance ($\Delta_\epsilon = 0$),  solutions of the equations \eqref{eom_tilde_u} or \eqref{eom_tilde_B} are functions of $\tau = -k\eta$ up to the coefficient $c_k$ introduced by the Bunch-Davies vacuum state. The analytic forms of the mode functions are given by
\begin{align}
	\tilde{u}_k &= c_k(\nu) \sqrt{\tau} H_\nu^{(1)}(\tau) &&= c_k(\nu) \tilde{B}_0(\tau) e^{i\tau}, \\
	\partial_\tau \tilde{u}_k &= \left(\frac{1}{2} - \nu\right)\frac{\tilde{u}_k}{\tau} + c_k(\nu) \sqrt{\tau} H_{\nu - 1}^{(1)}(\tau) 
	&&= c_k(\nu) \left(\partial_\tau \tilde{B}_0 + i \tilde{B}_0\right) e^{i\tau}.
\end{align}
Applying the expansion of the Hankel function in the early-time limit, where
\begin{align}
	\lim_{\tau\rightarrow\infty} H_{\nu}^{(1)}(\tau) = \sqrt{\frac{2}{\pi}} e^{-\frac{i}{2} \nu\pi} e^{-\frac{i}{4}\pi} \frac{e^{i\tau}}{\sqrt{\tau}},
\end{align}
we can match the above equations at the cutoff of the numerical computation $\tau = \tau_{\rm UV} \gg 1$ to get
\begin{align}
	\tilde{B}(\tau_{\rm UV}) = \sqrt{\frac{2}{\pi}} e^{-\frac{i}{2} \nu\pi} e^{-\frac{i}{4}\pi}, \quad 
	\left. \partial_\tau \tilde{B}_0 \right|_{\tau = \tau_{\rm UV}} = \left(\frac{1}{2} - \nu\right) \sqrt{\frac{2}{\pi}} e^{-\frac{i}{2} \nu\pi} e^{-\frac{i}{4}\pi} \frac{1}{\tau_{\rm UV}}.
\end{align}
It is also possible to solve the unscaled mode function $u_k = a^{-1}\tilde{u}_k$ with the definition $B_0(\tau) e^{i\tau} = \tau^{3/2} H_\nu^{(1)}(\tau)$. In this case the matching of initial conditions in the early-time limit gives
\begin{align}
	B_0(\tau_{\rm UV}) = \sqrt{\frac{2}{\pi}} e^{-\frac{i}{2} \nu\pi} e^{-\frac{i}{4}\pi} \tau_{\rm UV}, \quad 
	\left. \partial_\tau B_0 \right|_{\tau = \tau_{\rm UV}} = \left(\frac{3}{2} - \nu\right) \sqrt{\frac{2}{\pi}} e^{-\frac{i}{2} \nu\pi} e^{-\frac{i}{4}\pi}.
\end{align}
These initial conditions can apply to a general case with $\Delta_\epsilon > 0$ and they are also valid for the slow-roll scalaron solutions as given by the large $\tau$ limit of \eqref{eom_tilde_B_SR_phi}.

To deal with the fast oscillations of the mode functions in the early-time limit for the numerical integration over the time parameter $\tau$, it is convenient to adopt the Wick rotation method by taking $\tau \rightarrow i y$. Since in this work we do not have the analytic expression for the mode functions in $R^2$ inflation, we need to solve the mode functions by applying the Wick rotation to the equation of motion. For example, in the pivot-scale expansion approach we shall solve the rotated equation from \eqref{eom_tilde_B} as
\begin{align}
	-\frac{\partial^2}{\partial y^2} \tilde{B} + 2\frac{\partial}{\partial y} \tilde{B} 
	+ \left[\frac{M_\ast^2}{(iy)^{2+\Delta_\epsilon}}\left(\frac{k}{k_\ast}\right)^{\Delta_\epsilon} 
	+ \left(1+\frac{\Delta_\epsilon}{2}\right)\left(2+\frac{\Delta_\epsilon}{2}\right)\frac{1}{y^2}\right] \tilde{B} =0,
\end{align} 
with the rotated initial conditions
\begin{align}
	\tilde{B}(y_{\rm UV}) = \sqrt{\frac{2}{\pi}} e^{-\frac{i}{2} \nu\pi} e^{-\frac{i}{4}\pi}, \quad 
	\left. \partial_y \tilde{B}_0 \right|_{y = y_{\rm UV}} = \left(\frac{1}{2} - \nu\right) \sqrt{\frac{2}{\pi}} e^{-\frac{i}{2} \nu\pi} e^{-\frac{i}{4}\pi} \frac{1}{y_{\rm UV}}.
\end{align}
It is important to make sure that the oscillatory factor $e^{i(P+1)\tau} \rightarrow e^{-(P+1)y}$ in the integrand of \eqref{def_INT_PS} becomes a decaying function in the large $y$ limit.
See also the Appendix C of \cite{Chen:2009zp} for more details of the numerical integration under the Wick rotation.

\section{The simplest quantum primordial standard clock}\label{Appen_C}
In this section we provide the analytic formulas of the integrals \eqref{def_INT_PS} and \eqref{def_INT_bar} in the standard case with $\Delta_\epsilon = 0$, where $\tilde{B}(\tau) = \sqrt{\tau}H_\nu^{(1)}(\tau) e^{-i\tau}$ is the analytic solution for the factorized mode function of a massive scalar. The direct calculation of \eqref{def_INT_PS} gives
\begin{align}\label{def_ana_INT}
	I_{\rm NT}(0,\nu,P)=&\int_{0}^{\infty}  d\tau  \, e^{i P \tau} \tau^{-\frac{1}{2}} H_\nu^{(1)}(\tau) 
	\\\nonumber =&
	i(-iP)^{-\frac{1}{2}}
	\left\{ 
	(-2iP)^{-\nu} \left[\cot(\nu\pi) -i\right]  \Gamma\left(\frac{1}{2}+\nu\right) 
	\right.
	\\\nonumber &
	{}_2\tilde{F}_1\left[\frac{1+2\nu}{4},\frac{3+2\nu}{4}, 1 +\nu, \frac{1}{P^2}\right] -(-2iP)^{\nu} \csc(\nu\pi) \Gamma\left(\frac{1}{2}-\nu\right)
	\\\nonumber &
	\left.
	{}_2\tilde{F}_1\left[\frac{3-2\nu}{4},\frac{1-2\nu}{4}, 1-\nu, \frac{1}{P^2}\right]
	\right\},
\end{align}
and the result of \eqref{def_INT_bar} reads
\begin{align}\label{def_ana_INTbar}
	\bar{I}_{\rm NT}(0,\nu, P)=&\int_{0}^{\infty}  d\tau  \, e^{-i P \tau} \tau^{-\frac{1}{2}} H_\nu^{(2)}(\tau) 
	\\\nonumber =&
	-i(iP)^{-\frac{1}{2}}
	\left\{ 
	(2iP)^{-\nu} \left[\cot(\nu\pi) +i\right]  \Gamma\left(\frac{1}{2}+\nu\right) 
	\right.
	\\\nonumber &
	{}_2\tilde{F}_1\left[\frac{1+2\nu}{4},\frac{3+2\nu}{4}, 1+\nu, \frac{1}{P^2}\right] -(2iP)^{\nu} \csc(\nu\pi) \Gamma\left(\frac{1}{2}-\nu\right)
	\\\nonumber &
	\left.
	{}_2\tilde{F}_1\left[\frac{3-2\nu}{4},\frac{1-2\nu}{4}, 1-\nu, \frac{1}{P^2}\right]
	\right\},
\end{align}
where ${}_2\tilde{F}_1(a, b; c; z)\equiv {}_2F_1(a, b; c; z)/\Gamma(c)$ is the regularized hypergeometric function and $\tau = -k\eta$ is the dimensionless time parameter.
For $M_\ast > 3/2$, one replaces $\nu = i\mu$.

Taking the useful relations $\Gamma(\frac{1}{2}+\nu)\Gamma(\frac{1}{2}-\nu) = \pi/\cos(\pi\nu)$, $\Gamma(\nu)\Gamma(1-\nu) = \pi/\sin(\pi\nu)$ and applying the identical transformation of the hypergeometric function given in the Appendix F of \cite{Arkani-Hamed:2018kmz}, one can convert the above results to
\begin{align}\label{INT2_d3}
	\bar{I}_{\rm NT}(0, \nu, P) = i\frac{8\pi}{\sqrt{2}} \frac{e^{\frac{i}{2}\pi(\nu -1/2)}}{\cos(\pi\nu)} 
	{}_2F_1\left[\frac{1}{2}-\nu ,\frac{1}{2}+\nu, 1, \frac{1-P}{2}\right].
\end{align}
This is the familiar solution for the de Sitter boundary equations derived from the generator of special conformal transformations for two conformally coupled fields with one massive scalar (see Section 4.1 in \cite{Arkani-Hamed:2015bza}).

The integration $I_{\rm NT}$, $\bar{I}_{\rm NT}$ exhibit important singularities at $P = \pm 1$: 
The limit of $P \rightarrow -1$ means $k_1 + k_2 + k_3 \rightarrow 0$ which is nothing but the flat-space (high-energy) limit where bulk interactions are picked up at very early times ($\eta\rightarrow -\infty$). 

On the other hand, the limit of $P\rightarrow 1$ corresponds to the ``collapsed'' momentum configuration of the closed triangle ($k_1 + k_2 = k_3$). This singularity should be absent from the adiabatic vacuum condition and one can check that $I_{\rm NT}$, $\bar{I}_{\rm NT}$ is regular and continuous across $P =1$. Thus our numerical results are picked up at $P\rightarrow 1^\pm$. In fact, the singularity of collapsed triangle can be removed by the correct choices of coefficients in the most general solution obtained from the de Sitter four-point functions on the boundary \cite{Arkani-Hamed:2018kmz}. The three-point functions of slow-roll inflation (as investigated in this work) are reproduced from the soft limit for one of the external legs in the four-point correlator. This is manifest in the structure of $S_{\rm NT}$ given by \eqref{def_SNT23}, \eqref{def_SNT32} as a product of $I_{\rm NT}$ and $\bar{I}_{\rm NT}$ with one of the integration taken in the limit of $P \rightarrow 1$.


\begin{thebibliography}{99}
\bibitem{Planck:2013jfk}
P.~A.~R.~Ade \textit{et al.} [Planck],
``Planck 2013 results. XXII. Constraints on inflation,''
Astron. Astrophys. \textbf{571}, A22 (2014)
[arXiv:1303.5082 [astro-ph.CO]].
\bibitem{Planck:2015sxf}
P.~A.~R.~Ade \textit{et al.} [Planck],
``Planck 2015 results. XX. Constraints on inflation,''
Astron. Astrophys. \textbf{594}, A20 (2016)
[arXiv:1502.02114 [astro-ph.CO]].
\bibitem{Planck:2018jri}
Y.~Akrami \textit{et al.} [Planck],
``Planck 2018 results. X. Constraints on inflation,''
Astron. Astrophys. \textbf{641}, A10 (2020)
[arXiv:1807.06211 [astro-ph.CO]].

\bibitem{Starobinsky:1980te} 
A.~A.~Starobinsky,
``A New Type of Isotropic Cosmological Models Without Singularity,''
Phys.\ Lett.\  {\bf 91B}, 99 (1980).

\bibitem{SimonsObservatory:2018koc}
P.~Ade \textit{et al.} [Simons Observatory],
``The Simons Observatory: Science goals and forecasts,''
JCAP \textbf{02}, 056 (2019)
[arXiv:1808.07445 [astro-ph.CO]].
\bibitem{LiteBIRD:2020khw}
M.~Hazumi \textit{et al.} [LiteBIRD],
``LiteBIRD: JAXA's new strategic L-class mission for all-sky surveys of cosmic microwave background polarization,''
Proc. SPIE Int. Soc. Opt. Eng. \textbf{11443}, 114432F (2020)
[arXiv:2101.12449 [astro-ph.IM]].
\bibitem{CMB-S4:2016ple}
K.~N.~Abazajian \textit{et al.} [CMB-S4],
``CMB-S4 Science Book, First Edition,''
[arXiv:1610.02743 [astro-ph.CO]].

\bibitem{Braglia:2022ftm}
M.~Braglia, X.~Chen, D.~K.~Hazra and L.~Pinol,
``Back to the features: assessing the discriminating power of future CMB missions on inflationary models,''
JCAP \textbf{03}, 014 (2023)
[arXiv:2210.07028 [astro-ph.CO]].

\bibitem{Arkani-Hamed:2015bza}
N.~Arkani-Hamed and J.~Maldacena,
``Cosmological Collider Physics,''
[arXiv:1503.08043 [hep-th]].
\bibitem{Dimastrogiovanni:2015pla}
E.~Dimastrogiovanni, M.~Fasiello and M.~Kamionkowski,
``Imprints of Massive Primordial Fields on Large-Scale Structure,''
JCAP {\bf 1602}, 017 (2016).
[arXiv:1504.05993 [astro-ph.CO]].
\bibitem{Schmidt:2015xka}
F.~Schmidt, N.~E.~Chisari and C.~Dvorkin,
``Imprint of inflation on galaxy shape correlations,''
JCAP {\bf 1510}, no. 10, 032 (2015).
[arXiv:1506.02671 [astro-ph.CO]].
\bibitem{Kehagias:2015jha} 
A.~Kehagias and A.~Riotto,
``High Energy Physics Signatures from Inflation and Conformal Symmetry of de Sitter,''
Fortsch.\ Phys.\  {\bf 63}, 531 (2015)
[arXiv:1501.03515 [hep-th]].
\bibitem{Meerburg:2016zdz} 
P.~D.~Meerburg, M.~Münchmeyer, J.~B.~Muñoz and X.~Chen,
``Prospects for Cosmological Collider Physics,''
JCAP {\bf 1703}, no. 03, 050 (2017)
[arXiv:1610.06559 [astro-ph.CO]].
\bibitem{Lee:2016vti} 
H.~Lee, D.~Baumann and G.~L.~Pimentel,
``Non-Gaussianity as a Particle Detector,''
JHEP {\bf 1612}, 040 (2016)
[arXiv:1607.03735 [hep-th]].
\bibitem{Chen:2016uwp} 
X.~Chen, Y.~Wang and Z.~Z.~Xianyu,
``Standard Model Background of the Cosmological Collider,''
Phys.\ Rev.\ Lett.\  {\bf 118}, no. 26, 261302 (2017)
[arXiv:1610.06597 [hep-th]].
\bibitem{Chen:2016hrz} 
X.~Chen, Y.~Wang and Z.~Z.~Xianyu,
``Standard Model Mass Spectrum in Inflationary Universe,''
JHEP {\bf 1704}, 058 (2017)
[arXiv:1612.08122 [hep-th]].
\bibitem{Kehagias:2017cym} 
A.~Kehagias and A.~Riotto,
``On the Inflationary Perturbations of Massive Higher-Spin Fields,''
JCAP {\bf 1707}, no. 07, 046 (2017)
[arXiv:1705.05834 [hep-th]].
\bibitem{Franciolini:2017ktv} 
G.~Franciolini, A.~Kehagias and A.~Riotto,
``Imprints of Spinning Particles on Primordial Cosmological Perturbations,''
JCAP {\bf 1802}, no. 02, 023 (2018)
[arXiv:1712.06626 [hep-th]].
\bibitem{Kumar:2017ecc} 
S.~Kumar and R.~Sundrum,
``Heavy-Lifting of Gauge Theories By Cosmic Inflation,''
JHEP {\bf 1805}, 011 (2018)
[arXiv:1711.03988 [hep-ph]].
\bibitem{MoradinezhadDizgah:2018ssw} 
A.~Moradinezhad Dizgah, H.~Lee, J.~B.~Munoz and C.~Dvorkin,
``Galaxy Bispectrum from Massive Spinning Particles,''
arXiv:1801.07265 [astro-ph.CO].
\bibitem{Saito:2018omt} 
R.~Saito and T.~Kubota,
``Heavy Particle Signatures in Cosmological Correlation Functions with Tensor Modes,''
JCAP {\bf 1806}, no. 06, 009 (2018)
[arXiv:1804.06974 [hep-th]].
\bibitem{Kumar:2018jxz} 
S.~Kumar and R.~Sundrum,
``Seeing Higher-Dimensional Grand Unification In Primordial Non-Gaussianities,''
arXiv:1811.11200 [hep-ph].
\bibitem{Goon:2018fyu} 
G.~Goon, K.~Hinterbichler, A.~Joyce and M.~Trodden,
``Shapes of gravity: Tensor non-Gaussianity and massive spin-2 fields,''
arXiv:1812.07571 [hep-th].
\bibitem{Wu:2018lmx}
Y.~P.~Wu,
``Higgs as heavy-lifted physics during inflation,''
JHEP \textbf{04}, 125 (2019)
[arXiv:1812.10654 [hep-ph]].
\bibitem{Lu:2019tjj}
S.~Lu, Y.~Wang and Z.~Z.~Xianyu,
``A Cosmological Higgs Collider,''
JHEP \textbf{02}, 011 (2020)
[arXiv:1907.07390 [hep-th]].
\bibitem{Liu:2019fag}
T.~Liu, X.~Tong, Y.~Wang and Z.~Z.~Xianyu,
``Probing P and CP Violations on the Cosmological Collider,''
JHEP \textbf{04}, 189 (2020)
[arXiv:1909.01819 [hep-ph]].
\bibitem{Wang:2019gbi}
L.~T.~Wang and Z.~Z.~Xianyu,
``In Search of Large Signals at the Cosmological Collider,''
JHEP \textbf{02}, 044 (2020)
[arXiv:1910.12876 [hep-ph]].
\bibitem{Wang:2020ioa}
L.~T.~Wang and Z.~Z.~Xianyu,
``Gauge Boson Signals at the Cosmological Collider,''
JHEP \textbf{11}, 082 (2020)
[arXiv:2004.02887 [hep-ph]].
\bibitem{Bodas:2020yho}
A.~Bodas, S.~Kumar and R.~Sundrum,
``The Scalar Chemical Potential in Cosmological Collider Physics,''
JHEP \textbf{02}, 079 (2021)
[arXiv:2010.04727 [hep-ph]].
\bibitem{Aoki:2020zbj}
S.~Aoki and M.~Yamaguchi,
``Disentangling mass spectra of multiple fields in cosmological collider,''
JHEP \textbf{04}, 127 (2021)
[arXiv:2012.13667 [hep-th]].
\bibitem{Lu:2021wxu}
Q.~Lu, M.~Reece and Z.~Z.~Xianyu,
``Missing scalars at the cosmological collider,''
JHEP \textbf{12}, 098 (2021)
[arXiv:2108.11385 [hep-ph]].
\bibitem{Cui:2021iie}
Y.~Cui and Z.~Z.~Xianyu,
``Probing Leptogenesis with the Cosmological Collider,''
Phys. Rev. Lett. \textbf{129}, no.11, 111301 (2022)
[arXiv:2112.10793 [hep-ph]].
\bibitem{Pinol:2021aun}
L.~Pinol, S.~Aoki, S.~Renaux-Petel and M.~Yamaguchi,
``Inflationary flavor oscillations and the cosmic spectroscopy,''
Phys. Rev. D \textbf{107}, no.2, L021301 (2023)
[arXiv:2112.05710 [hep-th]].
\bibitem{Tong:2022cdz}
X.~Tong and Z.~Z.~Xianyu,
``Large Spin-2 Signals at the Cosmological Collider,''
[arXiv:2203.06349 [hep-ph]].
\bibitem{Chen:2022vzh}
X.~Chen, R.~Ebadi and S.~Kumar,
``Classical cosmological collider physics and primordial features,''
JCAP \textbf{08}, 083 (2022)
[arXiv:2205.01107 [hep-ph]].
\bibitem{Reece:2022soh}
M.~Reece, L.~T.~Wang and Z.~Z.~Xianyu,
``Large-field inflation and the cosmological collider,''
Phys. Rev. D \textbf{107}, no.10, L101304 (2023)
[arXiv:2204.11869 [hep-ph]].
\bibitem{Qin:2022lva}
Z.~Qin and Z.~Z.~Xianyu,
``Phase Information in Cosmological Collider Signals,''
[arXiv:2205.01692 [hep-th]].
\bibitem{Werth:2023pfl}
D.~Werth, L.~Pinol and S.~Renaux-Petel,
``Cosmological Flow of Primordial Correlators,''
[arXiv:2302.00655 [hep-th]].
\bibitem{Aoki:2023tjm}
S.~Aoki,
``Continuous spectrum on cosmological collider,''
JCAP \textbf{04}, 002 (2023)
[arXiv:2301.07920 [hep-th]].
\bibitem{Jazayeri:2023xcj}
S.~Jazayeri, S.~Renaux-Petel and D.~Werth,
``Shapes of the cosmological low-speed collider,''
JCAP \textbf{12}, 035 (2023)
[arXiv:2307.01751 [hep-th]].
\bibitem{Yin:2023jlv}
Y.~Yin,
``Cosmological collider signal from non-Bunch-Davies initial states,''
Phys. Rev. D \textbf{109}, no.4, 043535 (2024)
[arXiv:2309.05244 [hep-ph]].
\bibitem{Ema:2023dxm}
Y.~Ema and S.~Verner,
``Cosmological collider signatures of Higgs-R$^{2}$ inflation,''
JCAP \textbf{04}, 039 (2024)
[arXiv:2309.10841 [hep-ph]].
\bibitem{Pinol:2023oux}
L.~Pinol, S.~Renaux-Petel and D.~Werth,
``The Cosmological Flow: A Systematic Approach to Primordial Correlators,''
[arXiv:2312.06559 [astro-ph.CO]].
\bibitem{McCulloch:2024hiz}
C.~McCulloch, E.~Pajer and X.~Tong,
``A Cosmological Tachyon Collider: Enhancing the Long-Short Scale Coupling,''
[arXiv:2401.11009 [hep-th]].
\bibitem{Craig:2024qgy}
N.~Craig, S.~Kumar and A.~McCune,
``An Effective Cosmological Collider,''
[arXiv:2401.10976 [hep-ph]].
\bibitem{Werth:2024aui}
D.~Werth, L.~Pinol and S.~Renaux-Petel,
``CosmoFlow: Python Package for Cosmological Correlators,''
[arXiv:2402.03693 [astro-ph.CO]].
\bibitem{Cabass:2024wob}
G.~Cabass, O.~H.~E.~Philcox, M.~M.~Ivanov, K.~Akitsu, S.~F.~Chen, M.~Simonovi\'c and M.~Zaldarriaga,
``BOSS Constraints on Massive Particles during Inflation: The Cosmological Collider in Action,''
[arXiv:2404.01894 [astro-ph.CO]].
\bibitem{Sohn:2024xzd}
W.~Sohn, D.~G.~Wang, J.~R.~Fergusson and E.~P.~S.~Shellard,
``Searching for Cosmological Collider in the Planck CMB Data,''
[arXiv:2404.07203 [astro-ph.CO]].
\bibitem{Aoki:2024uyi}
S.~Aoki, L.~Pinol, F.~Sano, M.~Yamaguchi and Y.~Zhu,
``Cosmological Correlators with Double Massive Exchanges: Bootstrap Equation and Phenomenology,''
[arXiv:2404.09547 [hep-th]].


\bibitem{Chen:2009we}
X.~Chen and Y.~Wang,
``Large non-Gaussianities with Intermediate Shapes from Quasi-Single Field Inflation,''
Phys. Rev. D \textbf{81}, 063511 (2010)
[arXiv:0909.0496 [astro-ph.CO]].
\bibitem{Chen:2009zp}
X.~Chen and Y.~Wang,
``Quasi-Single Field Inflation and Non-Gaussianities,''
JCAP \textbf{04}, 027 (2010)
[arXiv:0911.3380 [hep-th]].
\bibitem{Baumann:2011nk} 
D.~Baumann and D.~Green,
``Signatures of Supersymmetry from the Early Universe,''
Phys.\ Rev.\ D {\bf 85}, 103520 (2012)
[arXiv:1109.0292 [hep-th]].
\bibitem{Chen:2012ge} 
X.~Chen and Y.~Wang,
``Quasi-Single Field Inflation with Large Mass,''
JCAP {\bf 1209}, 021 (2012)
[arXiv:1205.0160 [hep-th]].
\bibitem{Pi:2012gf} 
S.~Pi and M.~Sasaki,
``Curvature Perturbation Spectrum in Two-field Inflation with a Turning Trajectory,''
JCAP {\bf 1210}, 051 (2012)
[arXiv:1205.0161 [hep-th]].
\bibitem{McAllister:2012am}
L.~McAllister, S.~Renaux-Petel and G.~Xu,
``A Statistical Approach to Multifield Inflation: Many-field Perturbations Beyond Slow Roll,''
JCAP \textbf{10}, 046 (2012)
[arXiv:1207.0317 [astro-ph.CO]].
\bibitem{Assassi:2013gxa} 
V.~Assassi, D.~Baumann, D.~Green and L.~McAllister,
``Planck-Suppressed Operators,''
JCAP {\bf 1401}, 033 (2014)
[arXiv:1304.5226 [hep-th]].
\bibitem{Noumi:2012vr} 
T.~Noumi, M.~Yamaguchi and D.~Yokoyama,
``Effective field theory approach to quasi-single field inflation and effects of heavy fields,''
JHEP {\bf 1306}, 051 (2013)
[arXiv:1211.1624 [hep-th]].
\bibitem{Gong:2013sma}
J.~O.~Gong, S.~Pi and M.~Sasaki,
``Equilateral non-Gaussianity from heavy fields,''
JCAP {\bf 1311} (2013) 043
[arXiv:1306.3691 [hep-th]].
\bibitem{An:2017hlx} 
H.~An, M.~McAneny, A.~K.~Ridgway and M.~B.~Wise,
``Quasi Single Field Inflation in the non-perturbative regime,''
JHEP {\bf 1806}, 105 (2018)
[arXiv:1706.09971 [hep-ph]].
\bibitem{Tong:2017iat} 
X.~Tong, Y.~Wang and S.~Zhou,
``On the Effective Field Theory for Quasi-Single Field Inflation,''
JCAP {\bf 1711}, no. 11, 045 (2017)
[arXiv:1708.01709 [astro-ph.CO]].
\bibitem{Iyer:2017qzw} 
A.~V.~Iyer, S.~Pi, Y.~Wang, Z.~Wang and S.~Zhou,
``Strongly Coupled Quasi-Single Field Inflation,''
JCAP {\bf 1801}, no. 01, 041 (2018)
[arXiv:1710.03054 [hep-th]].
\bibitem{An:2017rwo} 
H.~An, M.~McAneny, A.~K.~Ridgway and M.~B.~Wise,
``Non-Gaussian Enhancements of Galactic Halo Correlations in Quasi-Single Field Inflation,''
Phys.\ Rev.\ D {\bf 97}, no. 12, 123528 (2018)
[arXiv:1711.02667 [hep-ph]].
\bibitem{Wang:2018tbf} 
Y.~Wang, Y.~P.~Wu, J.~Yokoyama and S.~Zhou,
``Hybrid Quasi-Single Field Inflation,''
JCAP {\bf 1807}, no. 07, 068 (2018)
[arXiv:1804.07541 [astro-ph.CO]].
\bibitem{Garcia-Saenz:2019njm}
S.~Garcia-Saenz, L.~Pinol and S.~Renaux-Petel,
``Revisiting non-Gaussianity in multifield inflation with curved field space,''
JHEP \textbf{01}, 073 (2020)
[arXiv:1907.10403 [hep-th]].


\bibitem{Arkani-Hamed:2018kmz}
N.~Arkani-Hamed, D.~Baumann, H.~Lee and G.~L.~Pimentel,
``The Cosmological Bootstrap: Inflationary Correlators from Symmetries and Singularities,''
JHEP \textbf{04}, 105 (2020)
[arXiv:1811.00024 [hep-th]].
\bibitem{Sleight:2019hfp}
C.~Sleight and M.~Taronna,
``Bootstrapping Inflationary Correlators in Mellin Space,''
JHEP \textbf{02}, 098 (2020)
[arXiv:1907.01143 [hep-th]].
\bibitem{Baumann:2019oyu}
D.~Baumann, C.~Duaso Pueyo, A.~Joyce, H.~Lee and G.~L.~Pimentel,
``The cosmological bootstrap: weight-shifting operators and scalar seeds,''
JHEP \textbf{12}, 204 (2020)
[arXiv:1910.14051 [hep-th]].
\bibitem{Baumann:2020dch}
D.~Baumann, C.~Duaso Pueyo, A.~Joyce, H.~Lee and G.~L.~Pimentel,
``The Cosmological Bootstrap: Spinning Correlators from Symmetries and Factorization,''
SciPost Phys. \textbf{11}, 071 (2021)
[arXiv:2005.04234 [hep-th]].
\bibitem{Goodhew:2020hob}
H.~Goodhew, S.~Jazayeri and E.~Pajer,
``The Cosmological Optical Theorem,''
JCAP \textbf{04}, 021 (2021)
[arXiv:2009.02898 [hep-th]].
\bibitem{Pajer:2020wxk}
E.~Pajer,
``Building a Boostless Bootstrap for the Bispectrum,''
JCAP \textbf{01}, 023 (2021)
[arXiv:2010.12818 [hep-th]].
\bibitem{Pimentel:2022fsc}
G.~L.~Pimentel and D.~G.~Wang,
``Boostless cosmological collider bootstrap,''
JHEP \textbf{10}, 177 (2022)
[arXiv:2205.00013 [hep-th]].
\bibitem{Jazayeri:2022kjy}
S.~Jazayeri and S.~Renaux-Petel,
``Cosmological bootstrap in slow motion,''
JHEP \textbf{12}, 137 (2022)
doi:10.1007/JHEP12(2022)137
[arXiv:2205.10340 [hep-th]].
\bibitem{Qin:2022fbv}
Z.~Qin and Z.~Z.~Xianyu,
``Helical inflation correlators: partial Mellin-Barnes and bootstrap equations,''
JHEP \textbf{04}, 059 (2023)
[arXiv:2208.13790 [hep-th]].
\bibitem{Xianyu:2022jwk}
Z.~Z.~Xianyu and H.~Zhang,
``Bootstrapping one-loop inflation correlators with the spectral decomposition,''
JHEP \textbf{04}, 103 (2023)
[arXiv:2211.03810 [hep-th]].
\bibitem{Wang:2022eop}
D.~G.~Wang, G.~L.~Pimentel and A.~Ach\'ucarro,
``Bootstrapping multi-field inflation: non-Gaussianities from light scalars revisited,''
JCAP \textbf{05}, 043 (2023)
[arXiv:2212.14035 [astro-ph.CO]].
\bibitem{Qin:2023ejc}
Z.~Qin and Z.~Z.~Xianyu,
``Closed-form formulae for inflation correlators,''
JHEP \textbf{07}, 001 (2023)
[arXiv:2301.07047 [hep-th]].
\bibitem{Arkani-Hamed:2023bsv}
N.~Arkani-Hamed, D.~Baumann, A.~Hillman, A.~Joyce, H.~Lee and G.~L.~Pimentel,
``Kinematic Flow and the Emergence of Time,''
[arXiv:2312.05300 [hep-th]].
\bibitem{Arkani-Hamed:2023kig}
N.~Arkani-Hamed, D.~Baumann, A.~Hillman, A.~Joyce, H.~Lee and G.~L.~Pimentel,
``Differential Equations for Cosmological Correlators,''
[arXiv:2312.05303 [hep-th]].

\bibitem{Antoniadis:2011ib}
I.~Antoniadis, P.~O.~Mazur and E.~Mottola,
``Conformal Invariance, Dark Energy, and CMB Non-Gaussianity,''
JCAP \textbf{09}, 024 (2012)
[arXiv:1103.4164 [gr-qc]].



\bibitem{Aoki:2020wzu}
S.~Aoki, T.~Noumi, F.~Sano and M.~Yamaguchi,
``Analytic formulae for inflationary correlators with dynamical mass,''
JHEP \textbf{24}, 073 (2020)
[arXiv:2312.09642 [hep-th]].


\bibitem{Sotiriou:2008rp}
T.~P.~Sotiriou and V.~Faraoni,
``f(R) Theories Of Gravity,''
Rev. Mod. Phys. \textbf{82}, 451-497 (2010)
[arXiv:0805.1726 [gr-qc]].
\bibitem{DeFelice:2010aj}
A.~De Felice and S.~Tsujikawa,
``f(R) theories,''
Living Rev. Rel. \textbf{13}, 3 (2010)
[arXiv:1002.4928 [gr-qc]].
\bibitem{Nojiri:2010wj}
S.~Nojiri and S.~D.~Odintsov,
``Unified cosmic history in modified gravity: from F(R) theory to Lorentz non-invariant models,''
Phys. Rept. \textbf{505}, 59-144 (2011)
[arXiv:1011.0544 [gr-qc]].
\bibitem{Nojiri:2017ncd}
S.~Nojiri, S.~D.~Odintsov and V.~K.~Oikonomou,
``Modified Gravity Theories on a Nutshell: Inflation, Bounce and Late-time Evolution,''
Phys. Rept. \textbf{692}, 1-104 (2017)
[arXiv:1705.11098 [gr-qc]].

\bibitem{Bezrukov:2007ep}
F.~L.~Bezrukov and M.~Shaposhnikov,
``The Standard Model Higgs boson as the inflaton,''
Phys. Lett. B \textbf{659}, 703-706 (2008)
[arXiv:0710.3755 [hep-th]].


\bibitem{He:2018gyf}
M.~He, A.~A.~Starobinsky and J.~Yokoyama,
``Inflation in the mixed Higgs-$R^2$ model,''
JCAP \textbf{05}, 064 (2018)
[arXiv:1804.00409 [astro-ph.CO]].











\bibitem{Chen:2015lza}
X.~Chen, M.~H.~Namjoo and Y.~Wang,
``Quantum Primordial Standard Clocks,''
JCAP \textbf{02}, 013 (2016)
[arXiv:1509.03930 [astro-ph.CO]].
\bibitem{Domenech:2018bnf}
G.~Dom\`enech, J.~Rubio and J.~Wons,
``Mimicking features in alternatives to inflation with interacting spectator fields,''
Phys. Lett. B \textbf{790}, 263-269 (2019)
[arXiv:1811.08224 [astro-ph.CO]].
\bibitem{Domenech:2020qay}
G.~Dom\`enech, X.~Chen, M.~Kamionkowski and A.~Loeb,
``Planck residuals anomaly as a fingerprint of alternative scenarios to inflation,''
JCAP \textbf{10}, 005 (2020)
[arXiv:2005.08998 [astro-ph.CO]].


\bibitem{Chen:2015dga}
X.~Chen, M.~H.~Namjoo and Y.~Wang,
``On the equation-of-motion versus in-in approach in cosmological perturbation theory,''
JCAP \textbf{01}, 022 (2016)
[arXiv:1505.03955 [astro-ph.CO]].





\bibitem{Weinberg:2005vy} 
S.~Weinberg,
``Quantum contributions to cosmological correlations,''
Phys.\ Rev.\ D {\bf 72}, 043514 (2005)
[hep-th/0506236].

\bibitem{Chen:2010xka} 
X.~Chen,
``Primordial Non-Gaussianities from Inflation Models,''
Adv.\ Astron.\  {\bf 2010}, 638979 (2010)
[arXiv:1002.1416 [astro-ph.CO]].





\bibitem{BICEP:2021xfz}
P.~A.~R.~Ade \textit{et al.} [BICEP and Keck],
``Improved Constraints on Primordial Gravitational Waves using Planck, WMAP, and BICEP/Keck Observations through the 2018 Observing Season,''
Phys. Rev. Lett. \textbf{127}, no.15, 151301 (2021)
[arXiv:2110.00483 [astro-ph.CO]].


\bibitem{Maggiore:2018sht}
M.~Maggiore,
``Gravitational Waves. Vol. 2: Astrophysics and Cosmology,''
Oxford University Press, 2018.

































\end{thebibliography}
\end{document}